\documentclass[aps,preprint]{revtex4}%
\usepackage{amsfonts}
\usepackage{amsmath}
\usepackage{amssymb}
\usepackage{amsthm}
\usepackage{graphicx}%
\usepackage{amsmath,amssymb}
\usepackage{enumitem}  
\usepackage{hyperref}
\usepackage[usenames, dvipsnames]{color} 
\usepackage{graphicx} 
\usepackage{comment}
\usepackage[normalem]{ulem}
\usepackage[utf8]{inputenc}
\usepackage{url}
 \usepackage{xcolor}
 \usepackage{multirow}
 \usepackage{amssymb}
\usepackage{pifont}
\providecommand{\U}[1]{\protect\rule{.1in}{.1in}}

\begin{document}
\title{Near-horizon behavior of nonequatorial accelerated~particle~motion and high
energy particle collisions}
\author{H. V. Ovcharenko}
\affiliation{Department of Physics and Technology, Kharkiv V.N. Karazin National
University, 4 Svoboda Square, Kharkov 61022, Ukraine}
\affiliation{Institute of Theoretical Physics, Faculty of Mathematics and Physics, Charles
University, Prague, V Holesovickach 2, 180 00 Praha 8, Czech Republic}
\author{O. B. Zaslavskii}
\affiliation{Department of Physics and Technology, Kharkiv V.N. Karazin National
University, 4 Svoboda Square, Kharkov 61022, Ukraine}

\begin{abstract}
We consider motion of a particle in the background of a stationary axially
symmetric generic black hole. A particle experiences the action of a force of
unspecified nature. We require the force to remain finite in a comoving frame.
The result is expressed in terms of several integers characterizing the Taylor
expansion of the metric coefficients near the horizon. We show that the polar
component of the four-velocity remains finite. As a result, the scenarios of
high-energy particle collisions, found previously in the context of the BSW
effect for equatorial motion, do not change qualitatively in the nonequatorial
case. The fact that the polar component is finite, also enables us to fill
some gaps in the description of the BSW effect in previous works. This
includes (i) classification of particle trajectories for non-equatorial
motion, (ii) more coherent derivation of kinematics of the BSW effect, (iii)
discussion of bands in which the BSW effect takes place. Our results have a
quite general character and can be used not only in description of high energy
particle collisions but also in diverse astrophysical problems for which
motion is not constrained by the equatorial plane.

\end{abstract}

\pacs{}
\maketitle
\tableofcontents

\section{Introduction}

Motion of particles in the background of black holes is a general and very
vast subject. Its properties are important both in astrophysics and
theoretical physics, especially for quite generic motion when particles are
not constrained to the equatorial plane. In doing so, there is important
question: which general constraints arise due to the presence of a force when
a particle moves near the horizon? The answer should take into account diverse
metric behavior for different types of the horizons and requirement of the
finiteness of such a force there. Among all possible applications of this
there is the one that is being discussed in literature intensively during the
last decade. This is the so-called Ba\~{n}ados-Silk-West effect (BSW after the
names of its authors) \cite{ban}. The main motivation of the present article
stems just from this subject. However, we would like to stress that the
results have more general area of applicability since we scrutiny interplay
between the force, kind of a trajectory and black hole metric.

According to the findings of \cite{ban}, if two particles move towards a black
hole and collide near it, the energy $E_{c.m.}$ in the center of mass frame
can grow without bound, provided one of particles has fine-tuned energy and
angular momentum. Originally, this result was obtained for (i) the Kerr black
hole, (ii) extremal horizon, (iii) free particles, (iv) motion within the
equatorial plane. Further, each of the first three condition was relaxed,
showing the BSW effect exists for equatorial motion around different types of
horizons, including nonextremal black holes \cite{gp} and generic
\textquotedblleft dirty\textquotedblright\ black holes \cite{prd}. As far as
the latter condition is concerned, a scenario where motion is nonequatorial
was analyzed in \cite{kd} for the Kerr black hole. In \cite{jh} it was
extended to generic black holes surrounded by matter (so-called dirty ones).

The goal of the present paper is to consider the most general type of scenario
when (i) a black hole is generic stationary axially symmetric one, (ii) the
horizon is quite generic, (iii) particles experience the influence of some
finite force of unspecified nature and (iv) move along nonequatorial
trajectories. Similar analysis was made in \cite{gen,gen2} where points (i) -
(iii) were taken into account. Now, we make a next step and consider also
point (iv). In doing so, we concentrate on the properties of a trajectory,
mainly on the behavior of the component of particle velocity due to the change
of a polar angle. In principle, one could expect the potential appearance of
new qualitative features of collision scenarios just to behavior of polar
angle. However, previous studies showed that for the Kerr metric this is not
the case. Below, we argue that also in general, account for nonequatorial
motion does not change properties of the BSW effect qualitatively. To
elucidate, whether or not nonequatorial motion brings such features in the set
of possible scenarios and affects the manifestation of the BSW effect,
analysis suggested in our paper is necessary. Processes with particles moving
arbitrarily near a black hole are relevant in astrophysics (see for example
the recent analysis of what an a near-horizon orbiter can see on the sky
\cite{del1}). And, what seems to be even more important, including
nonequatorial motion enables us to make descripton of the BSW effect as
general as possible.

The paper is organized as follows. In Sec. \ref{set} we list main formulas
describing the metric, the choice of observers acceleration in the comoving
frame. In Sec. \ref{near} we list the near-horizon expansion for the metric
coefficients needed in what follows. In Section \ref{sec_conds} we elucidate
the conditions when a force acting on a particle remains finite in the horizon
limit for different kinds of trajectory. In Sec. \ref{cons} we show how the
finiteness of the polar component of the four-velocity enables us to fill some
gaps in previous works on the BSW effect. In Sec. \ref{boost} we discuss the
relation between different frames relevant for our consideration. General
discussion of these results is given in Sec. \ref{concl}. In Appendix
\ref{comov} we explain how the comoving frame can be obtained from the OZAMO
frame. In Appendix \ref{ap} we suggest description of the frame transformation
from the group viewpoint in terms of rotation angles and parameters of a local
Lorentz transformation.

\section{General setup and description of non-geodesic particles\label{set}}

\subsection{Metric and particle motion}

We consider axially symmetric spacetimes
\begin{equation}
ds^{2}=-N^{2}dt^{2}+g_{\varphi\varphi}(d\varphi-\omega dt)^{2}+\dfrac{dr^{2}%
}{A}+g_{\theta\theta}d\theta^{2}. \label{metr}%
\end{equation}

Our convention in the choice of coordinates is $x^{0}=t,~x^{1}=r,~x^{2}%
=\theta,~x^{3}=\varphi$. In what follows, we will also use, for shortness,
notations $g_{\varphi}=g_{\varphi\varphi}$ and $g_{\theta}=g_{\theta\theta}$.

Let us consider motion of a particle in this background. If it moves freely,
it follows from the equations of motion that its four-velocity is equal to%
\begin{equation}
u^{\mu}=\left(  \frac{X}{N^{2}},\sigma\frac{\sqrt{A}}{N}P,u^{\theta},\frac
{L}{g_{\varphi}}+\omega\frac{X}{N^{2}}\right)  , \label{u}%
\end{equation}
where $u^{\theta}$ is an azimuthal velocity, $\sigma=\pm1,$ $X=E-\omega L$,
where $E$ and $L$ are the energy and angular momentum respectively, the
function $P$ is defined from the condition $u_{\mu}u^{\mu}=-1$ that gives us%
\begin{equation}
P^{2}=X^{2}-N^{2}\left(  1+\frac{L^{2}}{g_{\varphi}}+g_{\theta}(u^{\theta
})^{2}\right)  .
\end{equation}

Here, the quantities $E$ and $L$ have, accordingly, the meaning of energy and angular
momentum. They retain this meaning even under the presence of a
force. For free fall, they are conserved since the metric does not depend on
$t$ and $\varphi$. If force is present, $E$ and $L$ become coordinate-dependent functions.

\subsection{Frames and tetrads}

Acceleration of a particle is given by
\begin{equation}
a^{\mu}=u^{\nu}\nabla_{\nu}u^{\mu}=u^{\nu}\partial_{\nu}u^{\mu}+\Gamma
_{\nu\rho}^{\mu}u^{\nu}u^{\rho}.
\end{equation}

For our future computations it is sufficient to know the tetrad components of
acceleration $a_{C}^{(a)}=a^{\mu}\widetilde{e}_{\mu}^{(a)}$ in the comoving
frame $\widetilde{e}_{\mu}^{(a)}$ (CO for short). However, this cannot be done
from the very beginning. First of all, we start from the so-called
zero-angular momentum observer (ZAMO) frame \cite{72} and make transformations
to the CO frame, as is described in Appendix A. ZAMO frame is a direct
generalization of the static one in the case when the metric itself is also
static. It leads to simplification in calculations and is powerfull tool in
relativistic astrophysics \cite{72}. \cite{membrane}. There are two different
kinds of such a system. The first one is orbital ZAMO frame (OZAMO for
shortness), when a reference particle moves along the trajectory with
$r=const$. In general, it is not geodesics. Such a frame can be realized by
observers that carry a tetrad
\begin{align}
e_{(0)}^{\mu}&=\dfrac{1}{N}(1,0,0,\omega),~~~~~~~~~e_{(1)}^{\mu}=\sqrt{A}(0,1,0,0),
\label{ozamo1}\\
e_{(2)}^{\mu}&=\dfrac{1}{\sqrt{g_{\theta\theta}}}(0,0,1,0),~~~~~e_{(3)}^{\mu
}=\dfrac{1}{\sqrt{g_{\phi\phi}}}(0,0,0,1)\text{.} \label{ozamo2}%
\end{align}
We use letter \textquotedblleft O\textquotedblright\ to indicate that such an
observer orbits a black hole.

In this tetrad frame, the four-velocity reads%
\begin{equation}
u^{(a)}=\left(  \frac{X}{N},\sigma\frac{P}{N},\sqrt{g_{\theta}}u^{\theta
},\frac{L}{\sqrt{g_{\varphi}}}\right)  \text{.} \label{utet}%
\end{equation}

The second kind is the so-called FZAMO frame realized by free-falling
particles with $L=0$. But we will not analyze this frame until Sec. VI.

After transformations from the OZAMO frame to the CO one, we have, according
to the results of Appendix A, components of acceleration in the CO frame:%
\begin{align}
a_{C}^{(0)}  &  =0,\label{af_0}\\
a_{C}^{(1)}  &  =P\frac{\sqrt{A}}{N}\frac{\partial_{r}X+L\partial_{r}\omega
}{\sqrt{X^{2}-N^{2}}}-u^{\theta}\frac{\partial_{\theta}X+L\partial_{\theta
}\omega}{\sqrt{X^{2}-N^{2}}},\label{af_1}\\
a_{C}^{(2)}  &  =-\frac{1}{2}\frac{P^{\prime}}{P}\frac{A}{\sqrt{g_{\theta}}%
}\partial_{\theta}\left(  \frac{P^{2}}{AN^{2}}\right)  +\frac{1}%
{\sqrt{g_{\theta}}}\frac{P}{P^{\prime}}\left(  \frac{X}{N^{2}}(\partial
_{\theta}X+L\partial_{\theta}\omega)-\frac{L\partial_{\theta}L}{g_{\varphi}%
}\right)  +\nonumber\\
&  +\sqrt{g_{\theta}}u^{\theta}\frac{\sqrt{A}N}{P^{\prime}}\left(  \frac
{X}{N^{2}}(\partial_{r}X+L\partial_{r}\omega)-\frac{L\partial_{r}L}%
{g_{\varphi}}\right)  -P^{\prime}\frac{\sqrt{A}}{N}\frac{\partial
_{r}(g_{\theta}u^{\theta})}{\sqrt{g_{\theta}}},\label{af_2}\\
a_{C}^{(3)}  &  =\frac{1}{\sqrt{X^{2}-N^{2}}}\left\{  \frac{u^{\theta}}%
{\sqrt{g_{\varphi}}P^{\prime}}\left[  \left(  X^{2}-N^{2}\right)
\partial_{\theta}L-LX(\partial_{\theta}X+L\partial_{\theta}\omega)\right]
-\right. \nonumber\\
&  \left.  -\frac{P}{\sqrt{g_{\varphi}}P^{\prime}}\frac{\sqrt{A}}{N}\left[
(X^{2}-N^{2})\partial_{r}L-LX(\partial_{r}X+L\partial_{r}\omega)\right]
\right\}  . \label{af_3}%
\end{align}
here $P'$ is defined in (A11). Here, a new axis 1 is directed along a trajectory of a moving particle.

Our strategy consists in examining the behavior of each component of
acceleration near the horizon. In all cases, we do calculations in CO. In
doing so, we require each term in the components of acceleration to be regular
separately. In principle, one cannot exclude in advance the situations when
some terms diverge, but these divergences mutually cancel. However, we do not
consider such very special situations. Our approach includes also free motion
when acceleration is not only regular but equals zero exactly. We also have to
note that one may be interested if one can use other frames to investigate
forces with respect to them. This question is analyzed in Sec. \ref{boost}.

Although for our purposes we mainly need the CO frame, time to time we mention
the ZAMO one as well. In particular, the OZAMO frame is very convenient for
the kinematic analysis in terms of the three-dimensional velocities that
generalizes the results of \cite{k} (see below).

For the Kerr black hole, instructive analysis of relations between different
frames was done in \cite{pir3}). In the current work, we generalize it to a
generic dirty black hole.

\section{Near-horizon expansions\label{near}}

Now we are interested in near-horizon properties of the metric and particle
trajectories. Let us assume that the horizon is placed at $r=r_{h}.$ For this
hypersurface to be a horizon it is required that $A(r_{h})=N^{2}(r_{h})=0.$
This means that near the horizon we can write the Taylor expansions of the
functions $A$ and $N^{2}$ in the form
\begin{equation}
N^{2}=\kappa_{p}v^{p}+o(v^{p}),\text{ \ \ }A=A_{q}v^{q}+o(v^{q}),
\label{AN2_exp_gen}%
\end{equation}
where $v=r-r_{h},$ $p$ and $q$ are integers, and $\kappa_{p},$ $A_{q}$ are
coefficients that may depend on $\theta.$ Also, we assume that near the
horizon holds%
\begin{align}
\omega &  =\widehat{\omega}_{H}+\widehat{\omega}_{k}v^{k}+...+\widehat{\omega
}_{l-1}v^{l-1}+\omega_{l}v^{l}+o(v^{l}),\\
g_{a}  &  =g_{aH}+g_{am}v^{m}+o(v^{m}), \label{g_expan}%
\end{align}
where we use \textquotedblleft hat\textquotedblright\ notation for
coefficients that do not depend on $\theta.$ Here, $l,$ $k,$ $m$ are integers.
As was shown in \cite{ovzas23}, the regularity of the Ricci scalar (and other
scalar invariants) implies that integers $l$ and $k$ have to satisfy
\begin{equation}
k\geq\left[  \frac{p-q+3}{2}\right]  ,\text{ \ \ }l\geq\left[  \frac{p+1}%
{2}\right]  . \label{reg_conditions}%
\end{equation}

If these requirements are satified, then the corresponding hypersurface
$r_{h}$ is indeed a horizon (and not a singularity).

Also we assume that near the horizon expansions hold%
\begin{align}
X  &  =\widehat{X}_{0}+\widehat{X}_{s_{1}}v^{s_{1}}+...+\widehat{X}_{s_{2}%
-1}v^{s_{2}-1}+X_{s_{2}}v^{s_{2}}+o(v^{s_{2}}),\label{X_exp}\\
L  &  =\widehat{L}_{H}+\widehat{L}_{b_{1}}v^{b_{1}}+...+\widehat{L}_{b_{2}%
-1}v^{b_{2}-1}+L_{b_{2}}v^{b_{2}}+o(v^{b_{2}}),\label{L_exp}\\
u^{\theta}  &  =\widehat{u}_{c_{1}}^{\theta}u^{c_{1}}+..+\widehat{u}_{c_{2}%
-1}^{\theta}v^{c_{2}-1}+u_{c_{2}}^{\theta}v^{c_{2}}+o(v^{c_{2}}),
\label{uth_exp}%
\end{align}
where $s_{1},$ $s_{2},$ $b_{1},$ $b_{2},$ $c_{1}$ and $c_{2}$ are real numbers.

One reservation is in order. In principle, one can consider non-integers
$p,q,k,l$. However, this would poses a question about analytical continuaiton
of the metric across the horizon, properties of regularity of the metric, etc.
Even in the simplest case of static metric such issues are quite nontrivial
and require separate discussion (see \cite{kh} and references therein). In the
current work, we put this issue aside and make a simplest assumption that the
aforementioned number are integers.

Below, we examine the properties of different kinds of particles. Our strategy
is the following. We begin with kinematic consideration, then we consider
restrictions following the requirement of the finiteness of a force. In doing
so, we consider separately the cases of finite and infinite proper time needed
to reach the horizon.

\section{Conditions on near-horizon particles\label{sec_conds}}

\subsection{Usual particles\label{us}}

In what follows we examine the conditions when the tetrad components of
acceleration are finite. Let us at first consider usual particles.

\subsubsection{Kinematic considerations}

According to definition accepted in literature, for them $X\neq0$ on the
horizon and, hence, $\widehat{X}_{0}\neq0.$ Now let us consider quantity $P$
in this limit:%
\begin{equation}
P=\sqrt{X^{2}-N^{2}\left(  1+\frac{L^{2}}{g_{\varphi}}+g_{\theta}(u^{\theta
})^{2}\right)  }. \label{P_expr}%
\end{equation}

This quantity has to be real that requires $X^{2}\geq N^{2}\left(
1+\frac{L^{2}}{g_{\varphi}}+g_{\theta}(u^{\theta})^{2}\right)  .$ If
$u^{\theta}$ is finite (or even tend to zero) near the horizon, this condition
is automatically satisfied near the horizon because $N^{2}\rightarrow0,$ and
thus the whole expression $N^{2}\left(  1+\frac{L^{2}}{g_{\varphi}}+g_{\theta
}(u^{\theta})^{2}\right)  \rightarrow0,$ while $X^{2}\rightarrow$ $\widehat
{X}_{0}^{2}\geq0.$ However, if $u^{\theta}$ diverges near horizon, then this
condition becomes more complicated. Then, among three terms in $1+\frac{L^{2}%
}{g_{\varphi}}+g_{\theta}(u^{\theta})^{2}$ the term $g_{\theta}(u^{\theta
})^{2}$ is dominant that means that near the horizon%
\begin{equation}
N^{2}\left(  1+\frac{L^{2}}{g_{\varphi}}+g_{\theta}(u^{\theta})^{2}\right)
\sim v^{p+2c_{1}}.
\end{equation}

This quantity has to be finite (or tend to zero) for the quantity $P$ to be
real. This gives kinematic restriction on coefficient $c_{1}$%
\begin{equation}
c_{1}\geq-\frac{p}{2}. \label{c1_cond0}%
\end{equation}

If this is so, then in the near-horizon limit one has%
\begin{equation}
P=\left\{
\begin{array}
[c]{c}%
\widehat{X}_{0}\text{ if }c_{1}>-\frac{p}{2}\\
\sqrt{\widehat{X}_{0}^{2}-\kappa_{p}g_{\theta H}((u^{\theta})_{-p/2})^{2}%
}\text{ if }c_{1}=-\frac{p}{2}%
\end{array}
\right.  ,
\end{equation}
while in all cases $P^{\prime}=\widehat{X}_{0}$ in the horizon limit$.$

Before we proceed with the analysis of the forces, we have to investigate the
proper time of usual particles. As the radial component of the four-velocity
$u^{r}=\frac{dr}{d\tau},$ the proper time is given by%
\begin{equation}
\tau=\int\frac{dr}{u^{r}}=\sigma\int\frac{N}{\sqrt{A}}\frac{dr}{P}.
\label{tau}%
\end{equation}

As near the horizon for usual particles $P=O(1),$ we obtain%
\begin{equation}
\tau\sim\int v^{\frac{p-q}{2}}dr\sim\left\{
\begin{tabular}
[c]{l}%
$v^{\frac{p-q+2}{2}}$ if $q\neq p+2$\\
$\ln|v|$ if $q=p+2$%
\end{tabular}
\ \ \ \ \right.  . \label{tauus}%
\end{equation}

The proper time is regular if $q<p+2$, it diverges if $q\geq p+2.$ In the
first case such a particle can cross the horizon, in the second one it is
unable to do it.

\subsubsection{Dynamics considerations: regular proper time}

Let us start with the case when a usual particle can cross the horizon. As was
mentioned in the previous section, this happens if $q<p+2.$ We decided to
consider this case at first as it corresponds to the cases most relevant in
physics and astrophysics. For example, for the extremal Kerr spacetime (where
the BSW effect was originally found) $q=p$, and the condition $q<p+2$ is
automatically satisfied.

We remind the reader that we analyze the forces in the comoving frame. Let us
start with analysis of $a_{C}^{(1)}$ (\ref{af_1}). The first term will be
finite if $\partial_{r}X\sim\frac{N}{\sqrt{A}}$ and $\partial_{r}\omega
\sim\frac{N}{\sqrt{A}}.$ Integrating we find that these terms will be regular
if
\begin{equation}
s_{1}\geq\frac{p-q}{2}+1,\text{ \ \ }k\geq\frac{p-q}{2}+1. \label{s1_cond}%
\end{equation}

As in this subsection we assume $q<p+2,$ both corresponding terms tend to zero
for any non-negative $s_{1}$ and $k$.

The second term in $a_{C}^{(1)}$ (\ref{af_1}) will be regular if $u^{\theta
}\partial_{\theta}X$ and $u^{\theta}\partial_{\theta}\omega$ are regular (we
postpone analysis of these conditions).

Now let us move to $a_{C}^{(2)}$ (\ref{af_2}). We start with analysis of the
second term that will be finite if
\begin{equation}
\frac{X}{N^{2}}(\partial_{\theta}X+L\partial_{\theta}\omega)-\frac
{L\partial_{\theta}L}{g_{\varphi}}=O(1).
\end{equation}

This will be true if $\partial_{\theta}X\sim N^{2},$ $\partial_{\theta}%
\omega\sim N^{2}$ and $\partial_{\theta}L=O(1).$ This terms are regular if%
\begin{equation}
s_{2}\geq p,\text{ \ }l\geq p,\text{ \ }b_{2}\geq0. \label{s2_cond}%
\end{equation}

This allows us to analyze the second term in $a_{C}^{(1)}$ (\ref{af_1}) that
will be regular if $u^{\theta}\partial_{\theta}X$ and $u^{\theta}%
\partial_{\theta}\omega$ are regular. Let us start with $u^{\theta}%
\partial_{\theta}X$ that near the horizon behaves like%
\begin{equation}
u^{\theta}\partial_{\theta}X\sim v^{c_{1}+s_{2}}.
\end{equation}

However, as follows from (\ref{s2_cond}) and (\ref{c1_cond0}), the
corresponding degree $c_{1}+s_{2}\geq\frac{p}{2}$. This quantity is positive
and thus $u^{\theta}\partial_{\theta}X$ tends to zero. The same argument holds
for $u^{\theta}\partial_{\theta}\omega.$ Thus we have found that $a_{C}^{(1)}$
(\ref{af_1}) will be regular if both (\ref{c1_cond0}) and (\ref{s2_cond}) hold.

Now we are ready to analyze the first term in $a_{C}^{(2)}$ (\ref{af_2}). It
will be finite if
\begin{equation}
A\partial_{\theta}\left(  \frac{P^{2}}{AN^{2}}\right)  =A\partial_{\theta
}\left(  \frac{X^{2}}{AN^{2}}-\frac{1}{A}\left(  1+\frac{L^{2}}{g_{\varphi}%
}+g_{\theta}(u^{\theta})^{2}\right)  \right)  =O(1).
\end{equation}

In principle, this expression can be rewritten as%
\begin{equation}
\frac{\partial_{\theta}X^{2}}{N^{2}}-X^{2}\frac{\partial_{\theta}(AN^{2}%
)}{AN^{4}}+\frac{\partial_{\theta}A}{A}\left(  1+\frac{L^{2}}{g_{\varphi}%
}+g_{\theta}(u^{\theta})^{2}\right)  -\partial_{\theta}\left(  \frac{L^{2}%
}{g_{\varphi}}+g_{\theta}(u^{\theta})^{2}\right)  =O(1). \label{X_eq}%
\end{equation}

In (\ref{X_eq}) the first term is regular because of condition $s_{2}\geq p$
(see (\ref{s2_cond})). The second term in (\ref{X_eq}) will be finite if%
\begin{equation}
\partial_{\theta}(AN^{2})\sim AN^{4}.
\end{equation}

Equivalently, this condition may be written as%
\begin{equation}
\frac{\partial_{\theta}A}{A}+\frac{\partial_{\theta}N^{2}}{N^{2}}=O(N^{2}).
\label{AN2_cond}%
\end{equation}

It was analyzed and integrated in Appendix C in \cite{ovzas23}. In terms of
expansion coefficients, this condition can be written through such a
relation (eqs. (C5) and (C18) in \cite{ovzas23}):%
\begin{align}
A_{q}\kappa_{p}  &  =C_{p},\label{AN2_cond1}\\
\frac{A_{q+s}}{A_{q}}+\frac{\kappa_{p+s}}{\kappa_{p}}  &  =%
{\displaystyle\sum_{n=2}^{s}}
\frac{(-1)^{n}}{n}%
{\displaystyle\sum\limits_{k_{j}}}
\frac{n!}{k_{1}!...k_{n}!}\left[
{\displaystyle\prod\limits_{j=1}^{n}}
\left(  \frac{A_{q+j}}{A_{q}}\right)  ^{k_{j}}+%
{\displaystyle\prod\limits_{j=1}^{n}}
\left(  \frac{\kappa_{p+j}}{\kappa_{p}}\right)  ^{k_{j}}\right]  +C_{p+s}.
\label{AN2_cond2}%
\end{align}

Summation over $k_{j}$ is conducted over all combinantions of integers $k_{j}$
such that
\begin{equation}%
{\displaystyle\sum\limits_{j=1}^{n}}
jk_{j}=s.
\end{equation}

The coefficients $\{C_{p},C_{p+1},...,C_{2p}\}$ are constants, $s\subset
\{1,...,p\}.$

The fourth term in (\ref{X_eq}) is finite if $\partial_{\theta}L=O(1)$ and
$u^{\theta}\partial_{\theta}u^{\theta}=O(1).$ First condition is already
satisfied by (\ref{s2_cond}), while the second one is new and it gives us%
\begin{equation}
c_{1}+c_{2}\geq0. \label{c2_cond}%
\end{equation}

The third term in (\ref{X_eq}) is more complicated. To analyze this term we
have to note that $1+\frac{L^{2}}{g_{\varphi}}$ is finite near horizon, while
$g_{\theta}(u^{\theta})^{2}\sim v^{2c_{1}}.$ Thus the second term gives
$\partial_{\theta}A\sim v^{-2c_{1}+q}.$ We will investigate this condition later.

Now let us analyze the 4-th term in $a_{C}^{(2)}$ (\ref{af_2}). It will be
finite if
\begin{equation}
\frac{\sqrt{A}}{N}\frac{\partial_{r}(g_{\theta}u^{\theta})}{\sqrt{g_{\theta}}%
}=O(1),
\end{equation}

that requires
\begin{equation}
\partial_{r}(g_{\theta}u^{\theta})=\partial_{r}(g_{\theta})u^{\theta
}+g_{\theta}\partial_{r}(u^{\theta})=O\left(  \frac{N}{\sqrt{A}}\right)  .
\label{c1_cond}%
\end{equation}

The second term in (\ref{c1_cond}) is of order $v^{c_{1}-1}$. This term will
be regular if $c_{1}\geq\frac{p-q}{2}+1.$ The first term in (\ref{c1_cond}) is
of order $v^{c_{1}+m-1}.$ It will be regular if $c_{1}\geq\frac{p-q}{2}+1-m$
where $m$ was introduced in (\ref{g_expan})$.$ As $m$ is considered to be
non-negative (see (\ref{g_expan})), the condition coming from the finiteness
of the second term is stronger. Thus, the fourth term in $a_{C}^{(2)}$
(\ref{af_2}) will be regular if
\begin{equation}
c_{1}\geq\frac{p-q}{2}+1.
\end{equation}

As in this subsection we consider the case of finite proper time (that
requires $q<p+2$), this automatically means $c_{1}>0.$

Also note that for $c_{1}$ there is a condition (\ref{c1_cond0}). This
condition is less strict than $c_{1}\geq\frac{p-q}{2}+1$ for the case of a
finite proper time. Thus, we can combine both conditions and write
\begin{equation}
\text{If the proper time is finite (}q<p+2\text{), }c_{1}\geq\frac{p-q}{2}+1.
\label{c1_cond2}%
\end{equation}

The third term in $a_{C}^{(2)}$ (\ref{af_2}) is regular if%
\begin{equation}
u^{\theta}\frac{\sqrt{A}}{N}\partial_{r}X,\text{ \ \ }u^{\theta}\frac{\sqrt
{A}}{N}\partial_{r}\omega,\text{ \ \ }u^{\theta}N\sqrt{A}\partial_{r}L,
\end{equation}

are regular.

Substituting expansions for $A,$ $N^{2}$ and $u^{\theta},$ one finds that near
the horizon these expressions have order
\begin{equation}
v^{c_{1}-\frac{p-q}{2}+s_{1}-1},\text{ \ \ }v^{c_{1}-\frac{p-q}{2}+k-1},\text{
\ \ }v^{c_{1}+\frac{p+q}{2}+b_{1}-1},
\end{equation}
respectively.

However, as $c_{1}$ satisfies $c_{1}\geq\frac{p-q}{2}+1$ (see (\ref{c1_cond2}%
)), it is obvious that these terms are regular for any non-negative $s_{1},$
$k,$ $b_{1}.$

Now we are ready to find when the third term in (\ref{X_eq}) is regular. As
was already mentioned, regularity of the third term in (\ref{X_eq}) will be
achieved if $\frac{\partial_{\theta}A}{A}g_{\theta}(u^{\theta})^{2}$ is
regular. However, as we showed in (\ref{c1_cond2}), $c_{1}$ is positive, and
thus this term is automatically egular. Thus, we see that the only conditions
that have to hold for the functions $A$ and $N^{2}$ are (\ref{AN2_cond2}).

This finishes analysis of $a_{C}^{(2)}$ (\ref{af_2}). For regularity of this
term conditions (\ref{s2_cond}), (\ref{c2_cond}), (\ref{c1_cond2}) and
(\ref{AN2_cond2}) have to hold.

Now let us move to the $a_{C}^{(3)}$ (\ref{af_3}). We start with the second
line of $a_{C}^{(3)}$ (\ref{af_3}) that will be finite if
\begin{equation}
\frac{\sqrt{A}}{N}\partial_{r}L=O(1),\text{ }\frac{\sqrt{A}}{N}\partial
_{r}X=O(1),\text{ }\frac{\sqrt{A}}{N}\partial_{r}\omega=O(1).
\end{equation}

However, employing (\ref{s1_cond}), we see that the second and third
conditions are already satisfied, while the first one will be satisfied if%
\begin{equation}
b_{1}\geq\frac{p-q}{2}+1. \label{b1_cond}%
\end{equation}

The first term in $a_{C}^{(3)}$ (\ref{af_3}) will be finite if
\begin{equation}
u^{\theta}\partial_{\theta}L=O(1),\text{ \ \ }u^{\theta}\partial_{\theta
}X=O(1),\text{ \ \ }u^{\theta}\partial_{\theta}\omega=O(1).
\end{equation}

These conditions will be satisfied if
\begin{equation}
b_{2}\geq-c_{1},\text{ \ \ }s_{2}\geq-c_{1},\text{ \ \ }l\geq-c_{1}.
\label{b2_cond}%
\end{equation}

However, according to (\ref{c1_cond2}) and assuming $b_{2},s_{2}$ and $l$ to
be positive, conditions (\ref{b2_cond}) will be satified if
\begin{equation}
b_{2},s_{2},l\geq\frac{q-p-2}{2}. \label{b2_cond2}%
\end{equation}

But as in this subsection we consider only the case of finite proper time
($q<p+2$), these conditions are automatically satisfied by any non-negative
$b_{2},$ $s_{2}$ and $l.$

This completes the analysis of $a_{C}^{(3)}$ (\ref{af_3}). For this component
to be finite, conditions (\ref{b1_cond}) and (\ref{b2_cond2}) have to hold.

Now let us summarize all the conditions we obtained for the case of a regular
proper time ($q<p+2$). First of all, we have found that $s_{1}\geq\frac
{p-q}{2}+1$ (see (\ref{s1_cond})).

The quantity $s_{2}$ is constrained by (\ref{s2_cond}) and (\ref{b2_cond2})
and requirement $s_{2}>0$. But for a regular proper time ($q<p+2$) the number
$\frac{q-p-2}{2}$ in the condition (\ref{b2_cond2}) is negative, and the
condition (\ref{s2_cond}) (namely, $s_{2}\geq p$) appears to be the strongest
among these three.

The next integer is $k.$ It is constrained only by (\ref{s1_cond}).

The coefficient $l$ is constrained by (\ref{s2_cond}) and (\ref{b2_cond2}) and
requirement $l>0$. Situation with $l$ is the same as with $s_{2}$ and gives
$l\geq p.$

The coefficient $b_{1}$ is constrained only by (\ref{b1_cond}). Coefficient
$b_{2}$ is constrained by (\ref{s2_cond}), (\ref{b2_cond2}) and the condition
$b_{2}>0$. As we have already explained, the quantity $\frac{q-p-2}{2}$ is
negative for regular proper time. Thus the condition $b_{2}>0$ is more strict.

The last conditions constrain $c_{1}$ and $c_{2}.$ The coefficient $c_{1}$ is
constrained only by (\ref{c1_cond2})$.$ As was noted, for a regular proper
time (when $q<p+2$) this quantity is positive. As follows from (\ref{uth_exp}%
), $c_{2}\geq c_{1}.$ As $c_{1}$ is positive, we directly see that $c_{2}>0.$

The final list of conditions is%
\begin{equation}
\text{If proper time is regular (}q<p+2\text{), }s_{1},k,b_{1},c_{1}\geq
\frac{p-q}{2}+1,\text{ \ \ }s_{2},l\geq p,\text{ \ }b_{2},c_{2}>0,\text{ and
(\ref{AN2_cond2}).} \label{Usual_fin_tau_conds}%
\end{equation}

\subsubsection{Dynamics considerations: diverging proper time}

As was noted above, a usual particle will have diverging proper time if $q\geq
p+2.$ First of all, we note that, as we assume $p$ to be non-negative, this
automatically entails that $q>2.$ We remind the reader that we analyze the
forces in the comoving frame. At first we investigate the expression for
$a_{C}^{(1)}$ (\ref{af_1}). The first term is of order $v^{\frac{q-p}{2}%
+s_{1}-1}.$ However, as for diverging proper time $q\geq p+2,$ we see that
this term is regular for any positive $s_{1}.$ The same holds for the second
term that is of order $v^{\frac{q-p}{2}+k-1}.$ It will be regular for any
positive $k.$ The last two terms in $a_{C}^{(1)}$ (\ref{af_1}) we will analyze further.

Now let us analyze the $a_{C}^{(2)}$ (\ref{af_2}). We start with the second
term. Corresponding expressions are independent on the proper time and have
been already analyzed in the previous subsection. Corresponding conditions of
regularity of these expressions are given by (\ref{s2_cond}). As was also
noted in the previous section, conditions (\ref{s2_cond}) ensure that the last
two terms in $a_{C}^{(1)}$ (\ref{af_1}) are also regular. Finiteness of the
first term in $a_{C}^{(2)}$ (\ref{af_2}), as was argued in the previous
subsection, can be rewritten as (\ref{X_eq}). However, in this section we
slightly change the order in which we are analyzing corresponding terms.

The first term in (\ref{X_eq}) is regular if $s_{2}\geq p$ what is true even
for diverging proper time as the condition (\ref{s2_cond}) has to hold. Second
term in (\ref{X_eq}) is finite if conditions (\ref{AN2_cond1}-\ref{AN2_cond2})
hold. Now let us analyze the 4-th term in (\ref{X_eq}), namely $\partial
_{\theta}\left(  \frac{L^{2}}{g_{\varphi}}+g_{\theta}(u^{\theta})^{2}\right)
.$ Expanding the derivative, we obtain%
\begin{equation}
\partial_{\theta}\left(  \frac{L^{2}}{g_{\varphi}}+g_{\theta}(u^{\theta}%
)^{2}\right)  =\frac{\partial_{\theta}L^{2}}{g_{\varphi}}-\frac{L^{2}%
}{(g_{\varphi})^{2}}\partial_{\theta}g_{\varphi}+(\partial_{\theta}g_{\theta
})(u^{\theta})^{2}+g_{\theta}\partial_{\theta}(u^{\theta})^{2}.
\end{equation}

Here all four terms have to regular. But as the metric functions $g_{\theta},$
$g_{\varphi}$ (and all their derivatives) assumed to be finite, it directly
follows that both $L$ and $u^{\theta}$ (and their derivatives) have to be
regular. In terms of expansion coefficients (\ref{L_exp}-\ref{uth_exp}), this
means%
\begin{equation}
b_{1,2}\geq0,\text{ \ }c_{1,2}\geq0. \label{bc_us_diver}%
\end{equation}

The third term in (\ref{X_eq}), namely $\frac{\partial_{\theta}A}{A}\left(
1+\frac{L^{2}}{g_{\varphi}}+g_{\theta}(u^{\theta})^{2}\right)  ,$ is then
quite easy to analyze as $\frac{\partial_{\theta}A}{A}$ is regular as follows
from the expansion for $A$ (see (\ref{AN2_exp_gen})), while $\left(
1+\frac{L^{2}}{g_{\varphi}}+g_{\theta}(u^{\theta})^{2}\right)  ~$is regular
because of (\ref{bc_us_diver}). This finishes analysis of the (\ref{X_eq}),
and thus the first term in $a_{C}^{(2)}$ (\ref{af_2}).

Now we are left with the third and fourth terms in $a_{C}^{(2)}$ (\ref{af_2}).
Let us start with the fourth one, namely $P^{\prime}\frac{\sqrt{A}}{N}%
\frac{\partial_{r}(g_{\theta}u^{\theta})}{\sqrt{g_{\theta}}}$. This
expression, in fact, contains two terms $P^{\prime}\frac{\sqrt{A}}{N}%
\frac{g_{\theta}\partial_{r}(u^{\theta})}{\sqrt{g_{\theta}}}$ and $P^{\prime
}\frac{\sqrt{A}}{N}\frac{u^{\theta}\partial_{r}(g_{\theta})}{\sqrt{g_{\theta}%
}}$, the first is of order $v^{\frac{q-p}{2}+c_{1}-1},$ while the second is
$v^{\frac{q-p}{2}+c_{1}+m-1}.$ However, as we are considering the case of
diverging proper time ($q\geq p+2$) and as the condition $c_{1}\geq0$ has to
hold (\ref{bc_us_diver}), both these terms are regular (note that by definition $m\geq 1$).

The third term in $a_{C}^{(2)}$ (\ref{af_2}) contains three terms that are of
order of $v^{\frac{q-p}{2}+s_{1}-1},$ $v^{\frac{q-p}{2}+k-1}$ and
$v^{\frac{q+p}{2}+b_{1}-1}.$ However, as we are considering the case of
diverging proper time ($q\geq p+2$) and as, by definition $s_{1}$ and $k$ are
positive (along with $b_{1}$), we obtain that the corresponding terms are regular.

Thus we are only left with $a_{C}^{(3)}$ (\ref{af_3}). The first term, namely
$\frac{u^{\theta}}{\sqrt{X^{2}-N^{2}}\sqrt{g_{\varphi}}P^{\prime}}\left[
\left(  X^{2}-N^{2}\right)  \partial_{\theta}L-LX(\partial_{\theta}%
X+L\partial_{\theta}\omega)\right]  $ is, obviously, regular as all the
quantities $L,$ $X$ and $\omega$ are regular near the horizon. The second
term, namely $\frac{P}{\sqrt{X^{2}-N^{2}}\sqrt{g_{\varphi}}P^{\prime}}%
\frac{\sqrt{A}}{N}\left[  (X^{2}-N^{2})\partial_{r}L-LX(\partial
_{r}X+L\partial_{r}\omega)\right]  $ has three terms in it that are of order
of $v^{\frac{q-p}{2}+b_{1}-1}$, $v^{\frac{q-p}{2}+s_{1}-1},$ $v^{\frac{q-p}%
{2}+k-1}.$ All these three terms are regular for the diverging proper time
($q\geq p+2$) and positive $b_{1},$ $s_{1}$ and $k$

Let us now summarize all the conditions we have obtained so far for regular
forces. Coefficient $s_{1}$ unconstrained and can take any non-negative value,
$s_{2}$ is constrained by (\ref{s2_cond}). The coefficients $c_{1},$ $c_{2},$
$b_{1}$ and $b_{2}$ are given by (\ref{bc_us_diver}) and can take any
non-negative value$.$ Quantity $k$ is constrained only by the regularity
condition (\ref{reg_conditions}), $l$ is restricted by $l\geq p$
(\ref{s2_cond})$.$ The last requirement that has to hold is the set of
conditions (\ref{AN2_cond1}-\ref{AN2_cond2}). This can be summarized in such a
set of conditions%
\begin{equation}
s_{2}\geq p\text{, \ }l\geq p,\text{ }s_{1},c_{1,2},b_{1,2}\geq0\text{ and
(\ref{AN2_cond1}-\ref{AN2_cond2})}. \label{usual_diver_fin_conds}%
\end{equation}

The situation when $\tau\rightarrow\infty$ for usual particles refers to
so-called \textquotedblleft remote horizons\textquotedblright. For the
spherically symmetric metric the examples of such a kind were discussed in
\cite{prd08}. (In \cite{kh} such cases were classified as \textquotedblleft
null infinity\textquotedblright\ but we find such terminology in a given
context inaccurate. The lapse function $N\rightarrow0$, so the redshift grows
unbounded in this limit that is typical just of horizons.)

\subsection{Subcritical particles\label{sub}}

Now we are able to start an analysis of the subcritical particles. But before
we do this, we have to define precisely what it is meant by subcritical
particles. According to classification given in \cite{gen}, where authors
considered the case of equatorial motion, a subcritical particle is such that
near the horizon the quantity $X$ behaves according to
\begin{equation}
X=\widehat{X}_{s_{1}}v^{s_{1}}+o(v^{s_{1}}),
\end{equation}
where $0<s_{1}<\frac{p}{2}$. Note that as in \cite{gen} only equatorial motion
was investigated, corresponding expansion coefficients $\widehat{X}_{s_{1}}$
were inevitably independent on $\theta.$ In the case of non-equatorial motion,
generally it is not so. Thus, we have to assume an expansion for the quantity
$X$ in the form%
\begin{equation}
X=\widehat{X}_{s_{1}}v^{s_{1}}+...+\widehat{X}_{s_{2}-1}v^{s_{2}-1}+X_{s_{2}%
}v^{s_{2}}+o(v^{s_{2}}), \label{X_subcr_exp}%
\end{equation}
where $0<s_{1}<\frac{p}{2}$ and no additional constraint is imposed on
$s_{2}\geq s_{1}.$ Expansion of other parameters of the particle and metric
functions is assumed, as in the previous case, to be in the form%
\begin{align}
N^{2}  &  =\kappa_{p}v^{p}+o(v^{p}),\text{ \ \ }A=A_{q}v^{q}+o(v^{q}%
),\label{AN2_subcr_exp}\\
L  &  =\widehat{L}_{H}+\widehat{L}_{b_{1}}v^{b_{1}}+...+\widehat{L}_{b_{2}%
-1}v^{b_{2}-1}+L_{b_{2}}v^{b_{2}}+o(v^{b_{2}}),\\
u^{\theta}  &  =\widehat{u}_{c_{1}}^{\theta}v^{c_{1}}+..+\widehat{u}_{c_{2}%
-1}^{\theta}v^{c_{2}-1}+u_{c_{2}}^{\theta}v^{c_{2}}+o(v^{c_{2}}),\\
\omega &  =\widehat{\omega}_{H}+\widehat{\omega}_{k}v^{k}+...+\widehat{\omega
}_{l-1}v^{l-1}+\omega_{l}v^{l}+o(v^{l})\\
g_{a}  &  =g_{aH}+g_{am}v^{m}+o(v^{m}).
\end{align}

\subsubsection{Kinematical considerations}

Before we proceed with an analysis of acceleration components, we have to note
that several physical limitations come already from kinematics. Indeed, the
radial component of the four-velocity has to be real. This requires one to
have $P^{2}>0$:%
\begin{equation}
P^{2}=X^{2}-N^{2}\left(  1+\frac{L^{2}}{g_{\varphi}}+g_{\theta}(u^{\theta
})^{2}\right)  >0.
\end{equation}

The first term, namely $X^{2},$ is positive but tends to zero as $\left(
\widehat{X}_{s_{1}}\right)  ^{2}v^{2s_{1}}.$ The second and the third terms,
namely $-N^{2}\left(  1+\frac{L^{2}}{g_{\varphi}}\right)  $ are negative but
they tend to zero with the rate $v^{p}.$ As for the subcritical particles we
assumed that $s_{1}<p/2,$ these terms are of higher order then $X^{2}$ and do
not play crucial role in the near-horizon behavior of $P^{2}.$

However, the fourth term $-N^{2}g_{\theta}(u^{\theta})^{2}$ may have an
influence on the sign of $P^{2}.$ Indeed, this term is negative and near
horizon behaves as $-\kappa_{p}g_{\theta H}\left(  \widehat{u}_{c_{1}}%
^{\theta}\right)  ^{2}v^{2c_{1}+p}.$ Thus in the function $P^{2}$ there are
two potentially dominant terms: $\left(  \widehat{X}_{s_{1}}\right)
^{2}v^{2s_{1}}$ and $-\kappa_{p}g_{\theta H}\left(  \widehat{u}_{c_{1}%
}^{\theta}\right)  ^{2}v^{2c_{1}+p}.$ Depending on what term is dominant, we
obtain such various cases:

\begin{itemize}
\item Case $2s_{1}=2c_{1}+p.$ Then, both these terms are of the same order,
and $P^{2}$ near the horizon%
\begin{equation}
P^{2}\approx\left[  \left(  \widehat{X}_{s_{1}}\right)  ^{2}-\kappa
_{p}g_{\theta H}\left(  \widehat{u}_{c_{1}}^{\theta}\right)  ^{2}\right]
v^{2s_{1}}. \label{p_subcr_1}%
\end{equation}

The corresponding coefficient will be positive or at least nonnegative if
$\left(  \widehat{X}_{s_{1}}\right)  ^{2}\geq\kappa_{p}g_{\theta H}\left(
\widehat{u}_{c_{1}}^{\theta}\right)  ^{2}.$ Otherwise subcritical particle is
unable to reach the horizon.

\item Case $2s_{1}<2c_{1}+p.$ Then, the term $\left(  \widehat{X}_{s_{1}%
}\right)  ^{2}v^{2s_{1}}$ is dominant and we are able to write%
\begin{equation}
P^{2}\approx\left(  \widehat{X}_{s_{1}}\right)  ^{2}v^{2s_{1}}.
\label{p_subcr_2}%
\end{equation}

\item Case $2s_{1}>2c_{1}+p$ is impossible because in this case $P^{2}$ is
negative near the horizon.
\end{itemize}

Thus we have obtained that for subcritical particles kinematics requires%
\begin{equation}
0<s_{1}<\frac{p}{2},\text{ \ \ }c_{1}\geq s_{1}-\frac{p}{2}. \label{c1_subcr}%
\end{equation}

These conditions will be used further.

\bigskip

Now we are able to move to the analysis of forces acting on such particles. But before we do this, let us investigate the proper time required for such a particle to cross the horizon. The radial component of four-velocity in the leading order%
\begin{equation}
u^{r}=\sigma\frac{\sqrt{A}}{N}P\sim v^{\frac{q-p}{2}+s_{1}}.
\end{equation}

The proper time, required to cross the horizon, in the main order is given by%
\begin{equation}
\tau=\int\frac{dr}{u^{r}}\sim\left\{
\begin{tabular}
[c]{l}%
$v^{\frac{p-q+2}{2}-s_{1}}$ if $s_{1}\neq\frac{p-q+2}{2}$\\
$\ln|v|$ if $s_{1}=\frac{p-q+2}{2}$%
\end{tabular}
\ \ \ \right.  . \label{prop_time_subcr}%
\end{equation}

This quantity may either diverge (if $s_{1}\geq\frac{p-q+2}{2}$) or to be
regular (if $s_{1}<\frac{p-q+2}{2}$). These cases differ physically because if
the proper time diverges ($s_{1}\geq\frac{p-q+2}{2}$), this means that the
corresponding particle cannot cross the horizon. Meanwhile, if the proper time
is regular (this happens if $s_{1}<\frac{p-q+2}{2}$), such a particle can
cross the horizon.

\subsubsection{Dynamics considerations: diverging proper time}

In this section, we start with the case of diverging proper time. Combining
the defining condition for subcritical particles (\ref{c1_subcr}) and the
condition when the proper time (\ref{prop_time_subcr}) diverges, one obtains
that in this case%
\begin{equation}
\left\{
\begin{tabular}
[c]{l}%
$0<s_{1}<\frac{p}{2}$ if $q\geq p+2$\\
$\frac{p-q+2}{2}\leq s_{1}<\frac{p}{2}$ if $2<q<p+2$\\
Impossible if $q\leq2$%
\end{tabular}
\ \ \ \ \ \right.  . \label{subcr_s1_div}%
\end{equation}
As is explained above, we have to investigate components of acceleration in
the CO frame (\ref{af_0}-\ref{af_3}). We start our analysis with the
$a_{C}^{(1)}$ (\ref{af_1}) component.
\begin{equation}
a_{C}^{(1)}=P\frac{\sqrt{A}}{N}\frac{\partial_{r}X+L\partial_{r}\omega}%
{\sqrt{X^{2}-N^{2}}}-u^{\theta}\frac{\partial_{\theta}X+L\partial_{\theta
}\omega}{\sqrt{X^{2}-N^{2}}}.
\end{equation}

Let us, as previously, consider each term separately. First of all, from previous analysis of
kinematics (eqs. (\ref{p_subcr_1}) and (\ref{p_subcr_2})) it follows that the
function $P$ in this case is of order $\sim u^{s_{1}}.$ Thus the first term in
$a_{C}^{(1)}$ (\ref{af_1}) is of order%
\begin{equation}
P\frac{\sqrt{A}}{N}\frac{\partial_{r}X}{\sqrt{X^{2}-N^{2}}}\sim v^{s_{1}%
+\frac{q-p}{2}-1}.
\end{equation}

This term will be regular if $s_{1}+\frac{q-p}{2}-1\geq0,$ or, equivalently,
$s_{1}\geq\frac{p-q+2}{2}.$ Note that this condition has to be compatible with
(\ref{subcr_s1_div}). But looking at this condition you can notice that
(\ref{subcr_s1_div}) is stronger. Thus, the first term in (\ref{af_1}) will be
regular if (\ref{subcr_s1_div}) holds.

Now let us investigate the second term in $a_{C}^{(1)}$ (\ref{af_1}). This
term near the horizon behaves like%
\begin{equation}
P\frac{\sqrt{A}}{N}\frac{L\partial_{r}\omega}{\sqrt{X^{2}-N^{2}}}\sim
v^{\frac{q-p}{2}+k-1}.
\end{equation}

This term is regular if the regularity conditions hold (\ref{reg_conditions}).

The third term in $a_{C}^{(1)}$ (\ref{af_1}) is $u^{\theta}\frac
{\partial_{\theta}X}{\sqrt{X^{2}-N^{2}}}\sim v^{c_{1}+s_{2}-s_{1}}$. This term
is regular if
\begin{equation}
c_{1}\geq s_{1}-s_{2}. \label{c2_subcr_cond_1}%
\end{equation}

The last term in $a_{C}^{(1)}$ (\ref{af_1}) is $u^{\theta}\frac{L\partial
_{\theta}\omega}{\sqrt{X^{2}-N^{2}}}\sim v^{c_{1}+l-s_{1}}.$ This term will be
regular if%
\begin{equation}
c_{1}\geq s_{1}-l. \label{c2_subcr_cond_2}%
\end{equation}

Combining (\ref{c2_subcr_cond_2}) with (\ref{c2_subcr_cond_1}) one obtains%
\begin{equation}
c_{1}\geq s_{1}-\min(l,s_{2}). \label{c1_subcr_cond3}%
\end{equation}

Now we are able to investigate the $a_{C}^{(2)}$ (\ref{af_2}). The first term
is%
\begin{equation}
-\frac{1}{2}\frac{P^{\prime}}{P}\frac{A}{\sqrt{g_{\theta}}}\partial_{\theta
}\left(  \frac{P^{2}}{AN^{2}}\right)  .
\end{equation}

Note that as we are considering the case of subcritical particles, both
quantities $P$ and $P^{\prime}$ tend to zero near horizon with the same rate
$\sim v^{s_{1}}.$ Thus, expanding the expression for $P^{2}$ one would obtain
that finiteness of this term will give exactly the condition (\ref{X_eq}),
namely%
\begin{equation}
\frac{\partial_{\theta}X^{2}}{N^{2}}-X^{2}\frac{\partial_{\theta}(AN^{2}%
)}{AN^{4}}+\frac{\partial_{\theta}A}{A}\left(  1+\frac{L^{2}}{g_{\varphi}%
}+g_{\theta}(u^{\theta})^{2}\right)  -\partial_{\theta}\left(  \frac{L^{2}%
}{g_{\varphi}}+g_{\theta}(u^{\theta})^{2}\right)  =O(1).
\end{equation}

The first term in this expression is $\frac{\partial_{\theta}X^{2}}{N^{2}%
}=\frac{2X\partial_{\theta}X}{N^{2}}\sim v^{s_{1}+s_{2}-p}.$ It will be
regular if%
\begin{equation}
s_{2}\geq p-s_{1}. \label{s2_subcr_diver_cond}%
\end{equation}

The last term $\partial_{\theta}\left(  \frac{L^{2}}{g_{\varphi}}+g_{\theta
}(u^{\theta})^{2}\right)  $ has been already analyzed in the previous section.
It was shown that for the regularity of this term quantities $L$ and
$u^{\theta}$ have to be regular and conditions (\ref{bc_us_diver}) have to
hold. This automatically leads to regularity of the third term. However, the
second term is not so easy. Finiteness of this term requires us to have
$\frac{\partial_{\theta}(AN^{2})}{AN^{2}}\sim v^{p-2s_{1}}.$ This condition
will be satisfied if the conditions (\ref{AN2_cond1}-\ref{AN2_cond2}) hold for
all $s\subset\{0,...,p-2s_{1}\}.$

Now let us analyze the second term in $a_{C}^{(2)}$ (\ref{af_2}), namely
$\frac{1}{\sqrt{g_{\theta}}}\frac{P}{P^{\prime}}\left(  \frac{X}{N^{2}%
}(\partial_{\theta}X+L\partial_{\theta}\omega)-\frac{L\partial_{\theta}%
L}{g_{\varphi}}\right)  .$ As for subcritical particles near the horizon
$P\sim P^{\prime}\sim v^{s_{1}},$ this term will be regular if $\frac
{X}{N^{2}}(\partial_{\theta}X+L\partial_{\theta}\omega)-\frac{L\partial
_{\theta}L}{g_{\varphi}}$ is regular. The first term in this expression is of
order $v^{s_{1}+s_{2}-p}$ and will be regular if (\ref{s2_subcr_diver_cond})
holds. The second term is of order $v^{s_{1}+l-p}$ and it will be regular if
\begin{equation}
l\geq p-s_{1}. \label{l_subcr_diver_cond}%
\end{equation}

The last term is regular for non-negative $b_{2}$
(\ref{bc_us_diver})$.$

Now let us move to the third term in $a_{C}^{(2)}$ (\ref{af_2}), namely
$u^{\theta}\frac{\sqrt{A}N}{P^{\prime}}\left(  \frac{X}{N^{2}}(\partial
_{r}X+L\partial_{r}\omega)-\frac{L\partial_{r}L}{g_{\varphi}}\right)  .$ The
first term in this expression is of order $v^{c_{1}+\frac{q-p}{2}+s_{1}-1}.$
But it can be easily seen that if the conditions (\ref{subcr_s1_div}) and
(\ref{bc_us_diver}) hold, this term is regular. The second term here is of
order $v^{c_{1}+\frac{q-p}{2}+k-1}$ and will be regular if
(\ref{reg_conditions}) and (\ref{bc_us_diver}) hold. The last term in this
expression is of order $v^{c_{1}+\frac{q+p}{2}-s_{1}+b_{1}-1}.$ This term is
also regular because $c_{1},b_{1}\geq0$ according to (\ref{bc_us_diver}) and
as $0<s_{1}<p/2$ for subcritical particles and $q>2$ for diverging proper time.

The last term in $a_{C}^{(2)}$ (\ref{af_2}) is $P^{\prime}\frac{\sqrt{A}}%
{N}\frac{\partial_{r}(g_{\theta}u^{\theta})}{\sqrt{g_{\theta}}}=P^{\prime
}\frac{\sqrt{A}}{N}\frac{\partial_{r}(g_{\theta})u^{\theta}+g_{\theta}%
\partial_{r}(u^{\theta})}{\sqrt{g_{\theta}}}.$ As for subcritical particles
near the horizon holds $P^{\prime}\approx X\sim v^{s_{1}},$ the first term in
this expression is of order $v^{s_{1}+\frac{q-p}{2}+c_{1}},$ while the second
one is $v^{s_{1}+\frac{q-p}{2}+c_{1}-1}.$ However, both these terms are
regular because $c_{1}\geq0$ according to (\ref{bc_us_diver}) and
$\frac{p-q+2}{2}\leq s_{1}$ according to (\ref{subcr_s1_div}).

Now let us investigate the last component $a_{C}^{(3)}$ (\ref{af_3}). Note
that for the analysis of this component it is useful to remind that $P\approx
P^{\prime}\approx\sqrt{X^{2}-N^{2}}\approx X\sim v^{s_{1}}.$ The first term in
$a_{C}^{(3)}$ (\ref{af_3}) is of order $v^{c_{1}+b_{2}}.$ As $c_{1}\geq0$
(\ref{bc_us_diver}), we see that this term is regular if $b_{2}\geq0.$ The
second term in $a_{C}^{(3)}$ (\ref{af_3}) is of order $v^{c_{1}+s_{2}-s_{1}}.$
This term is regular because $c_{1}\geq0$ (\ref{bc_us_diver}) and as
$s_{2}\geq s_{1}$ by the definition$.$ The third term is of order
$v^{c_{1}-s_{1}+l}.$ This term is regular because $c_{1}\geq0$
(\ref{bc_us_diver}) and as $l\geq\left[  \frac{p+1}{2}\right]  $ because of
(\ref{reg_conditions}) and as $0<s_{1}<p/2$ by the definition$.$ The fourth
term in $a_{C}^{(3)}$ (\ref{af_3}) is of order $v^{s_{1}+\frac{q-p}{2}%
+b_{1}-1}.$ But it is regular because of conditions (\ref{subcr_s1_div}) and
(\ref{bc_us_diver}). The fifth term in $a_{C}^{(3)}$ (\ref{af_3}) is of order
$v^{\frac{q-p}{2}+s_{1}-1}$ and it is regular because of (\ref{subcr_s1_div}).
The last term in $a_{C}^{(3)}$ (\ref{af_3}) is of order $v^{\frac{q-p}{2}%
+k-1}$ and it is regular because of regularity conditions
(\ref{reg_conditions}).

Summarizing all the conditions, we see that $s_{1}$ is constrained only by
conditions (\ref{subcr_s1_div}), while $s_{2}$ is constrained by
(\ref{s2_subcr_diver_cond}). The coefficients $c_{1,2},$ $b_{1,2}$ are
constrained only by (\ref{bc_us_diver}). Coefficient $k$ is constrained only
by regularity conditions (\ref{reg_conditions}), while $l$ is constrained by
(\ref{l_subcr_diver_cond}). But in addition to these conditions
(\ref{AN2_cond1}-\ref{AN2_cond2}) have to hold for all $s\subset
\{0,...,p-2s_{1}\}$ . Thus for subcritical particles with diverging proper
time we require only
\begin{align}
c_{1,2}  &  \geq0,\text{ }b_{1,2}\geq0,\left\{
\begin{tabular}
[c]{l}%
$0<s_{1}<\frac{p}{2}$ if $q\geq p+2$\\
$\frac{p-q+2}{2}\leq s_{1}<\frac{p}{2}$ if $2<q<p+2$\\
Impossible if $q\leq2$%
\end{tabular}
\ \ \ \ \right.  ,\text{ }s_{2}\geq p-s_{1},\text{ }(\ref{reg_conditions}%
),\text{ }l\geq p-s_{1}\nonumber\\
&  \text{along with conditions (\ref{AN2_cond1}-\ref{AN2_cond2}) for }%
s\subset\{0,...,p-2s_{1}\}. \label{subcr_diver_fin_conds}%
\end{align}

\subsubsection{Dynamics considerations: regular proper time}

If proper time is regular for subcritical particles (what happens if
$s_{1}<\frac{p-q+2}{2}$, see (\ref{prop_time_subcr})), such a particle can
easily cross the horizon. As for subcritical particles the condition
$0<s_{1}<p/2$ has to hold, combining this with the condition $s_{1}%
<\frac{p-q+2}{2}$ one obtains that subcritical particles will have regular
proper time if%
\begin{equation}
\left\{
\begin{tabular}
[c]{l}%
Impossible if $q\geq p+2$\\
$0<s_{1}<\frac{p-q+2}{2}$ if $2<q<p+2$\\
$0<s_{1}<\frac{p}{2}$ if $q\leq2$%
\end{tabular}
\ \ \ \right.
\end{equation}

The first equation we are going to investigate is (\ref{af_1}). The first term
in $a_{C}^{(1)}$ is of order%
\begin{equation}
P\frac{\sqrt{A}}{N}\frac{\partial_{r}X}{\sqrt{X^{2}-N^{2}}}\sim u^{\frac
{q-p}{2}+s_{1}-1}.
\end{equation}

This term will be regular if $s_{1}\geq\frac{p-q+2}{2}.$ However, regularity
of the proper time requires one to have $s_{1}<\frac{p-q+2}{2},$ see
(\ref{prop_time_subcr}), thus without any further analysis we can say that
\textbf{it is impossible to have subcritical particles with regular proper
time. }

\subsection{Critical particles\label{crit}}

The next step is to consider an analog of the so-called critical particles. In
\cite{gen}, where only equatorial motion was investigated, critical particles
were defined in such a way that expansion for the quantity $X$ was given in
the form%
\begin{equation}
X=X_{p/2}v^{p/2}+o(v^{p/2}).
\end{equation}

In our case we generalize this definition and assume that for critical
particles the expansion holds%
\begin{equation}
X=\widehat{X}_{p/2}v^{p/2}+...+\widehat{X}_{s_{2}-1}v^{s_{2}-1}+X_{s_{2}%
}v^{s_{2}}+o(v^{s_{2}}), \label{X_expan_cr}%
\end{equation}
where $s_{2}\geq p/2.$ Here, as previously, hat means that the corresponding
quantity is independent of $\theta.$ Expansions for all other quantities, as
in the previous sections, are given by%
\begin{align}
N^{2}  &  =\kappa_{p}v^{p}+o(v^{p}),\text{ \ \ }A=A_{q}v^{q}+o(v^{q}),\\
L  &  =\widehat{L}_{H}+\widehat{L}_{b_{1}}v^{b_{1}}+...+\widehat{L}_{b_{2}%
-1}v^{b_{2}-1}+L_{b_{2}}v^{b_{2}}+o(v^{b_{2}}),\\
u^{\theta}  &  =\widehat{u}_{c_{1}}^{\theta}v^{c_{1}}+..+\widehat{u}_{c_{2}%
-1}^{\theta}v^{c_{2}-1}+u_{c_{2}}^{\theta}v^{c_{2}}+o(v^{c_{2}}),\\
\omega &  =\widehat{\omega}_{H}+\widehat{\omega}_{k}v^{k}+...+\widehat{\omega
}_{l-1}v^{l-1}+\omega_{l}v^{l}+o(v^{l}),\\
g_{a}  &  =g_{aH}+g_{am}v^{m}+o(v^{m}). \label{gm_expan}%
\end{align}

\subsubsection{Kinematical considerations}

We start with the analysis of the kinematics of such particles. Critical
particles may exist if the radial component of the four-velocity is real near
the horizon. This is equivalent to the requirement that $P^{2}$ is
non-negative. Namely, we require
\begin{equation}
P^{2}=X^{2}-N^{2}\left(  1+\frac{L^{2}}{g_{\varphi}}+g_{\theta}(u^{\theta
})^{2}\right)  \geq0. \label{Ppos}%
\end{equation}

Expanding this near $u=0$, one obtains%
\begin{equation}
P^{2}\approx\left[  \widehat{X}_{p/2}^{2}-\kappa_{p}\left(  1+\frac{L_{H}^{2}%
}{g_{\varphi H}}\right)  \right]  v^{p}-\kappa_{p}g_{\theta H}\left(
\widehat{u}_{c_{1}}^{\theta}\right)  ^{2}v^{p+2c_{1}}. \label{P2_cr}%
\end{equation}

There are two terms that may potentially be divergent, and we have to
investigate corresponding cases distinctly.

\begin{itemize}
\item Case $c_{1}<0.$ In this case the second term is dominant and
\begin{equation}
P^{2}=-\kappa_{p}g_{\theta H}\left(  \widehat{u}_{c_{1}}^{\theta}\right)
^{2}v^{p+2c_{1}}+o(v^{p+2c_{1}}).
\end{equation}

However, $P^{2}$ is negative and thus this case is prohibited. This means that
\textit{for critical particles, it is impossible to have divergent polar
velocity.}

\item Case $c_{1}=0.$ In this case both terms in (\ref{P2_cr}) are of the same
order, and
\begin{equation}
P^{2}=\left[  \widehat{X}_{p/2}^{2}-\kappa_{p}\left(  1+\frac{L_{H}^{2}%
}{g_{\varphi H}}+g_{\theta H}\left(  \widehat{u}_{0}^{\theta}\right)
^{2}\right)  \right]  v^{p}+o(v^{p}).
\end{equation}

This quantity will be non-negative if
\begin{equation}
\widehat{X}_{p/2}^{2}\geq\kappa_{p}\left(  1+\frac{L_{H}^{2}}{g_{\varphi H}%
}+g_{\theta H}\left(  \widehat{u}_{0}^{\theta}\right)  ^{2}\right)  .
\label{crit_ineq}%
\end{equation}

\item Case $c_{1}>0.$ In this case the first term in (\ref{P2_cr}) is of the
dominant order, and one can write%
\begin{equation}
P^{2}=\left[  \widehat{X}_{p/2}^{2}-\kappa_{p}\left(  1+\frac{L_{H}^{2}%
}{g_{\varphi H}}\right)  \right]  v^{p}+o(v^{p}).
\end{equation}

Such particles may reach the horizon if
\begin{equation}
\widehat{X}_{p/2}^{2}\geq\kappa_{p}\left(  1+\frac{L_{H}^{2}}{g_{\varphi H}%
}\right)  . \label{crit_ineq_2}%
\end{equation}

To summarize, we have to say that critical particles are possible only in the
case when $c_{1}\geq0.$
\end{itemize}

Before we analyze forces acting on critical particles, let us investigate
behavior of the proper time. Radial component of the four-velocity is given by%

\begin{equation}
u^{r}=\sigma\frac{\sqrt{A}}{N}P\sim v^{\frac{q}{2}}.
\end{equation}

The proper time required to cross the horizon, in the leading order is given
by%
\begin{equation}
\tau=\int\frac{dr}{u^{r}}\sim\left\{
\begin{tabular}
[c]{l}%
$v^{\frac{2-q}{2}}$ if $q\neq2$\\
$\ln |v|$ if $q=2$%
\end{tabular}
\ \ \ \right.  . \label{prop_time_cr}%
\end{equation}

The proper time diverges if $q\geq2$ (extremal and ultraextremal horizons)$.$
Otherwise ($q<2$) it is regular (non-extremal horizons). We will investigate
these cases distinctly.

\subsubsection{Dynamics considerations: diverging proper time ($q\geq2$)}

In this case we also use the expressions for acceleration in CO (\ref{af_0}%
-\ref{af_3}). Let us start with the radial component (\ref{af_1})%
\begin{equation}
a_{C}^{(1)}=P\frac{\sqrt{A}}{N}\frac{\partial_{r}X+L\partial_{r}\omega}%
{\sqrt{X^{2}-N^{2}}}-u^{\theta}\frac{\partial_{\theta}X+L\partial_{\theta
}\omega}{\sqrt{X^{2}-N^{2}}}.
\end{equation}

The first term is of order (here we used that for critical particles $s=p/2$)%
\begin{equation}
P\frac{\sqrt{A}}{N}\frac{\partial_{r}X}{\sqrt{X^{2}-N^{2}}}\sim v^{\frac{q}%
{2}-1}.
\end{equation}

As in this subsection we consider the case of diverging proper time
($q\geq2),$ we immediately see that this term is regular.

The second term in (\ref{af_1}) is of order%
\begin{equation}
P\frac{\sqrt{A}}{N}\frac{L\partial_{r}\omega}{\sqrt{X^{2}-N^{2}}}\sim
v^{\frac{q-p}{2}+k-1}.
\end{equation}

As the regularity conditions (\ref{reg_conditions}) have to hold, this term is
inevitably regular.

The third term in (\ref{af_1}) is of order%
\begin{equation}
\frac{u^{\theta}}{\sqrt{X^{2}-N^{2}}}\partial_{\theta}X\sim v^{c_{1}%
+s_{2}-p/2}.
\end{equation}

This term will be regular if
\[
s_{2}\geq p/2-c_{1}.
\]

Meanwhile, for critical particles we assumed that $s_{2}\geq p/2$ (see the
text after (\ref{X_expan_cr})), so this condition is automatically satisfied.

The last term in (\ref{af_1}) is of order%
\begin{equation}
\frac{u^{\theta}}{\sqrt{X^{2}-N^{2}}}L\partial_{\theta}\omega\sim
v^{c_{1}+l-p/2}.
\end{equation}

This term will be regular if
\begin{equation}
c_{1}\geq p/2-l.
\end{equation}

As from regularity conditions (\ref{reg_conditions}) follows that $l\geq p/2,$
this condition is satified for any $c_{1}\geq0$.%
\begin{equation}
c_{1}\geq p/2-\min(l,p/2). \label{c2_cond_cr}%
\end{equation}

The next component is (\ref{af_2}).%
\begin{align}
a_{C}^{(2)}  &  =-\frac{1}{2}\frac{P^{\prime}}{P}\frac{A}{\sqrt{g_{\theta}}%
}\partial_{\theta}\left(  \frac{P^{2}}{AN^{2}}\right)  +\frac{1}%
{\sqrt{g_{\theta}}}\frac{P}{P^{\prime}}\left(  \frac{X}{N^{2}}(\partial
_{\theta}X+L\partial_{\theta}\omega)-\frac{L\partial_{\theta}L}{g_{\varphi}%
}\right)  +\nonumber\\
&  +\sqrt{g_{\theta}}u^{\theta}\frac{\sqrt{A}N}{P^{\prime}}\left(  \frac
{X}{N^{2}}(\partial_{r}X+L\partial_{r}\omega)-\frac{L\partial_{r}L}%
{g_{\varphi}}\right)  -P^{\prime}\frac{\sqrt{A}}{N}\frac{\partial
_{r}(g_{\theta}u^{\theta})}{\sqrt{g_{\theta}}}.
\end{align}

Let us start with the terms in the second line. The first two terms there
(namely $\sqrt{g_{\theta}}u^{\theta}\frac{\sqrt{A}}{P^{\prime}}\frac{X}%
{N}\partial_{r}X$ and $\sqrt{g_{\theta}}u^{\theta}\frac{\sqrt{A}}{P^{\prime}%
}\frac{X}{N}L\partial_{r}\omega$) are of the same order as the first two terms
in (\ref{af_1}), and thus regular. The third term in the second line is of
order (here we used that for critical particles $P^{\prime}\sim N$)%
\begin{equation}
\frac{\sqrt{A}N}{P^{\prime}}\frac{L\partial_{r}L}{g_{\varphi}}\sim
v^{b_{1}+\frac{q}{2}-1}.
\end{equation}

This term is regular if $b_{1}\geq1-\frac{q}{2}.$ However, now we investigate
only particles with diverging proper time, meaning that $q\geq2.$ Thus the
right hand side of the condition $b_{1}\geq1-\frac{q}{2}$ is always negative
and thus the third term in the second line in (\ref{af_2}) will be regular for
any non-negative value of $b_{1}.$

The fourth term in the second line in (\ref{af_2}) is given by
\begin{equation}
P^{\prime}\frac{\sqrt{A}}{N}\frac{\partial_{r}(g_{\theta}u^{\theta})}%
{\sqrt{g_{\theta}}}=P^{\prime}\frac{\sqrt{A}}{N\sqrt{g_{\theta}}}(\partial
_{r}(g_{\theta})u^{\theta}+g_{\theta}\partial_{r}(u^{\theta})).
\end{equation}

The second term here is dominant, because the first one is of order $\sim
v^{\frac{q}{2}+m-1+c_{1}}$ (here $m$ is positive, see (\ref{gm_expan}))$,$
while the second is $\sim v^{\frac{q}{2}+c_{1}-1}.$ As for diverging proper
time we require $q\geq2$ and as $c_{1}\geq0$ (comes from kinematics), this
term is non-divergent. Thus, the whole second line in (\ref{af_2}) is regular.

Now let us analyze the first line in (\ref{af_2}). The first line therein is
fiven by%
\begin{equation}
-\frac{1}{2}\frac{P^{\prime}}{P}\frac{A}{\sqrt{g_{\theta}}}\partial_{\theta
}\left(  \frac{P^{2}}{AN^{2}}\right)  .
\end{equation}

Using that for critical particles both $P\sim N$ and $P^{\prime}\sim N,$ and
expanding expression for $P^{2}$ (see (\ref{P_expr})), we obtain that the
condition of finiteness of this term is exactly the (\ref{X_eq})
\begin{equation}
\frac{\partial_{\theta}X^{2}}{N^{2}}-X^{2}\frac{\partial_{\theta}(AN^{2}%
)}{AN^{4}}+\frac{\partial_{\theta}A}{A}\left(  1+\frac{L^{2}}{g_{\varphi}%
}+g_{\theta}(u^{\theta})^{2}\right)  -\partial_{\theta}\left(  \frac{L^{2}%
}{g_{\varphi}}+g_{\theta}(u^{\theta})^{2}\right)  =O(1).
\end{equation}

Analysis of this expression is somehow analogous to the case of divering
proper time for subcritical particles. The first term here is automatically
non-divergent as it is $\sim v^{s_{2}-p/2}.$ As $s_{2}\geq p/2$
(\ref{X_expan_cr}), this term is regular. The second term in (30) is automatically finite
as for critical particles $s_1=p/2$. Third and fourth terms in (\ref{X_eq}) are
regular because $u^{\theta}$ is regular (as the condition $c_{1}\geq0$ follows
from kinematics), thus both $\left(  \frac{L^{2}}{g_{\varphi}}+g_{\theta
}(u^{\theta})^{2}\right)  $ and $\left(  1+\frac{L^{2}}{g_{\varphi}}%
+g_{\theta}(u^{\theta})^{2}\right)  $ are regular.

In the first line in (\ref{af_2}) we are left with the $\frac{1}%
{\sqrt{g_{\theta}}}\frac{P}{P^{\prime}}\left(  \frac{X}{N^{2}}(\partial
_{\theta}X+L\partial_{\theta}\omega)-\frac{L\partial_{\theta}L}{g_{\varphi}%
}\right)  $ term. However, as for critical particles $X\sim N,$
$P^{\prime}\sim P$ this term is regular for all $c_{2},b_{2},l\geq0.$

Now let us move to the last component, (\ref{af_3}).%

\begin{align*}
a_{C}^{(3)}  &  =\frac{1}{\sqrt{X^{2}-N^{2}}}\left\{  \frac{u^{\theta}}%
{\sqrt{g_{\varphi}}P^{\prime}}\left[  \left(  X^{2}-N^{2}\right)
\partial_{\theta}L-LX(\partial_{\theta}X+L\partial_{\theta}\omega)\right]
-\right. \\
&  \left.  -\frac{P}{\sqrt{g_{\varphi}}P^{\prime}}\frac{\sqrt{A}}{N}\left[
(X^{2}-N^{2})\partial_{r}L-LX(\partial_{r}X+L\partial_{r}\omega)\right]
\right\}
\end{align*}
\qquad The first term $\frac{u^{\theta}}{\sqrt{g_{\varphi}}P^{\prime}}%
\sqrt{X^{2}-N^{2}}\partial_{\theta}L$ $\sim v^{b_{2}+c_1}$ and is regular for any
$b_{2}\geq0$ (as $c_1\geq 0$). The second term $\frac{u^{\theta}}{\sqrt{g_{\varphi}}P^{\prime}%
}\frac{LX}{\sqrt{X^{2}-N^{2}}}\partial_{\theta}X$ $\sim v^{s_{2}-p/2+c_1}$ and it
is regular as $s_{2}\geq p/2$ (\ref{X_expan_cr}). The third term is of order
of $\frac{u^{\theta}}{\sqrt{g_{\varphi}}P^{\prime}}\frac{L^{2}X}{\sqrt
{X^{2}-N^{2}}}\partial_{\theta}\omega\sim v^{l-p/2+c_1}$ and is regular because of
regularity condition (\ref{reg_conditions}). The fourth term is of order
$\frac{P}{\sqrt{g_{\varphi}}P^{\prime}}\frac{\sqrt{A}}{N}\sqrt{X^{2}-N^{2}%
}\partial_{r}L\sim v^{\frac{q}{2}-1+b_{1}}$ and it is regular because for
diverging proper time $q\geq2$ and as $b_{1}\geq0.$ The fifth term is of order
$\frac{1}{\sqrt{X^{2}-N^{2}}}\frac{P}{\sqrt{g_{\varphi}}P^{\prime}}\frac
{\sqrt{A}}{N}LX\partial_{r}X\sim v^{\frac{q}{2}-1}$ and it is regular for
diverging proper time $q\geq2.$ The sixth term is of order $\frac{1}%
{\sqrt{X^{2}-N^{2}}}\frac{P}{\sqrt{g_{\varphi}}P^{\prime}}\frac{\sqrt{A}}%
{N}L^{2}X\partial_{r}\omega\sim v^{\frac{q}{2}-\frac{p}{2}+k-1}$ and is
regular because of regularity condition (\ref{reg_conditions}).

Thus we see that if the proper time diverges (that happens in the case
$q\geq2$), the regularity of the forces acting on such particles does not
require anything new as compared to kinematic restrictions (namely $c_{1}%
\geq0,$ (\ref{crit_ineq}) and (\ref{crit_ineq_2})) except for conditions of
regularity of horizons (\ref{reg_conditions}).

\subsubsection{Dynamics considerations: regular proper time $(q<2)$}

We start with the (\ref{af_1}) component.%
\begin{equation}
a_{C}^{(1)}=P\frac{\sqrt{A}}{N}\frac{\partial_{r}X+L\partial_{r}\omega}%
{\sqrt{X^{2}-N^{2}}}-u^{\theta}\frac{\partial_{\theta}X+L\partial_{\theta
}\omega}{\sqrt{X^{2}-N^{2}}}.
\end{equation}

The first term is of order
\begin{equation}
P\frac{\sqrt{A}}{N}\frac{\partial_{r}X}{\sqrt{X^{2}-N^{2}}}\sim v^{\frac{q}%
{2}-1},
\end{equation}
where we took into account that for critical particles $X\sim N\sim P$.

However, in this subsection we assume that the proper time is regular (so that
$q<2$), and thus this term diverges. This means that such a situation is impossible.

To conclude this Section devoted to critical particles, we have to note that
they can reach the horizon only if $c_{1}\geq0$ and conditions
(\ref{crit_ineq}) and (\ref{crit_ineq_2}) have to hold. Forces acting on such
particles, are regular only if $q\geq2.$ However, this leads to the diverging
proper time (namely, $c_{1}\geq0,$ (\ref{crit_ineq}) and (\ref{crit_ineq_2})).

\subsection{Ultracritical particles\label{ultra}}

The last type of particles we are going to investigate are ultracritical ones.
In \cite{gen} this type of particles was defined in such a way that for them
an expansion holds%
\begin{equation}
X=X_{p/2}v^{p/2}+o(v^{p/2}).
\end{equation}

\subsubsection{Kinematic considerations}

We require that for ultracritical particles, the expansion coefficients are such that several dominant
terms in the expression for $P$ cancel each other and one obtains that the
radial component of the four-velocity behaves as $u^{r}\sim v^{i}$, where
$i>\frac{q}{2}$ (the latter value was obtained for critical particles)$.$ To
find what conditions have to hold in this case, let us consider the function
$P$:%
\begin{equation}
P^{2}=X^{2}-N^{2}\left(  1+\frac{L^{2}}{g_{\varphi}}+g_{\theta}(u^{\theta
})^{2}\right)  .
\end{equation}

As was noted in the previous section, $P^{2}$ will be non-negative if
$c_{1}\geq0.$ Then, $P$ becomes%
\begin{equation}
P^{2}=\left[  X_{p/2}^{2}-\kappa_{p}\left(  1+\frac{L_{H}^{2}}{g_{\varphi H}%
}+g_{\theta H}(u_{0}^{\theta})^{2}\right)  \right]  v^{p}+o(v^{p}).
\end{equation}

The radial component of the four-velocity is given by $u^{r}=\sigma\frac
{\sqrt{A}}{N}P.$ This quantity will be of order of $v^{i}$ if we choose
coefficients $X_{s}$ in such a way that%
\begin{equation}
X^{2}-N^{2}\left(  1+\frac{L^{2}}{g_{\varphi}}+g_{\theta}(u^{\theta}%
)^{2}\right)  \sim v^{2i+p-q}. \label{ultracr_cond}%
\end{equation}

We will not write explicit expression for $X_{s}$ because it is quite
complicated and it is unnecessary.

\subsubsection{Dynamics considerations}

Detailed analysis of forces is not required because this case differs from the
case of critical particles only in the behavior of the function $P$ with $P$
decreases to zero faster for ultracritical particles than for critical
particles. To see this, we have to note that in (\ref{af_0}-\ref{af_3}) the
function $P$ appears only in the numerator. In turn, this implies that the
conditions of the regularity of forces for critical particles lead to the fact
that for ultracritical particles these conditions also ensure the regularity
of forces.

This completes consideration of different types of particles. Meanwhile, we
must clarify one important feature. In general, expression (\ref{Ppos})
depends on the angle $\theta.$ This means that fulfilment of this condition
depends not only on parameters of a particle but also on angle $\theta.$ The
most representative in this sense is the situation near the poles of black
holes. For regular black holes (without any conical singularities) it holds
typically that near the poles $g_{\varphi H}\sim\theta^{2}\,\ $or $g_{\varphi
H}\sim(\pi-\theta)^{2}.\,$ Thus the right hand side of (\ref{crit_ineq})
behaves as $\sim\theta^{-2}$ as one approaches pole $\theta=0$, and the
inequality (\ref{Ppos}) cannot be satisfied everywhere. This creates the
\textquotedblleft belts\textquotedblright\ near equator where particle can
move and where different scenarios of collision may occur. (For the details of
these belts in the Kerr metric and their role for properties of critical
particles and, hence, the BSW effect see Sec. IV of \cite{kd}. However, we
would like to stress that the existence of belts is a generic property of
nonequatorial motion, inherent not only to critical particles.)

Concluding all the previous sections, we present Table I, which summarizes the
kinematics and regularity of forces constraints. It can be seen that the
regularity of forces requires that usual particles can have either regular or
diverging proper time, while fine-tuned particles can have only diverging
proper time.\begin{table}[ptb]
\centering%
\begin{tabular}
[c]{|c|c|c|c|}\hline
Type of particle & Kinematic restrictions & Regular proper time & Diverging
proper time\\\hline
Usual ($s_{1}=0$) & None & (\ref{Usual_fin_tau_conds}) &
(\ref{usual_diver_fin_conds})\\\hline
Subcr. ($0<s_{1}<p/2$) & (\ref{c1_subcr}) & Impossible &
(\ref{subcr_diver_fin_conds})\\\hline
Critical ($s_{1}=p/2$) & (\ref{crit_ineq}) if $c_{1}=0$ and (\ref{crit_ineq_2}%
) if $c_{1}>0$ & Impossible & Same as for k.r.\\\hline
Ultracr. ($s_{1}>p/2$) & (\ref{ultracr_cond}) and $c_{1}\geq0$ & Impossible &
Same as for k.r.\\\hline
\end{tabular}
\caption{List of conditions that have to hold for different types of
particles. The second column represents conditions coming from kinematic
conditions. The conditions in the last two colums are the ones that have to
hold in addition to the kinematic restrictions to have regular forces.
\textquotedblleft Same as for k.r.\textquotedblright\ means that the kinematic
restictions (denoted for shortness as \textquotedblleft k.r.\textquotedblright%
) are sufficient for the regularity of the force.}%
\end{table}

\section{Consequences of finiteness of $u^{\theta}$\label{cons}}

In previous Sections, we have shown that regularity of force implies that
negative $c_{1}$ is incompatible with the regularity of acceleration and,
hence, $u^{\theta}$ and the third component of the velocity in the OZAMO frame
(\ref{utet}) near the horizon remain finite (if $c_{1}=0$) or even tends to
zero (if $c_{1}>0$). This has important consequences. First of all, it means
that a general classification of possible scenarios of the BSW effect found in
the equatorial case remains valid here as well. Namely, the BSW effect occurs
for collisions of such particles: usual and subcritical, two subcritical,
usual or subcritical collides with critical or ultracritial (see Table II in
\cite{gen} for details).

Also, the fact that $c_{1}\geq0$ enables us to fill some gaps in previous
consideration. General explanation of the BSW effect was suggested earlier
\cite{cqg} in which the following expansion of the four-velocity was used:%
\begin{equation}
u^{\mu}=\frac{l^{\mu}}{2\alpha}+\beta N^{\mu}+s^{\mu}\text{,}%
\end{equation}
where $l^{\mu}$ and $N^{\mu}$ are null vectors and $s^{\mu}$ is the space-like
one that is orthogonal to $l^{\mu}$ and $N^{\mu}$. The same expansion was used
in \cite{bif} to elucidate the role of the bifurcation surface as a potential
particle accelerator. In both cases, the assumption was made according to
which $s_{\mu}s^{\mu}$ remained finite. Now this is proved since%
\begin{equation}
s_{\mu}s^{\mu}=g_{\phi}\left(  u^{\phi}\right)  ^{2}+g_{\theta}\left(
u^{\theta}\right)  ^{2}\text{,}%
\end{equation}
where both terms are finite.

The same assumption of the finiteness of $u^{\theta}$ was made in Appendix B
of \cite{jmp} to show that in the OZAMO frame a usual particle hits the
horizon perpendicularly. Now this assumption is proven.

\section{Three-velocities and relevant frames\label{boost}}

Kinematically, unbounded growth of the energy of collisions $E_{c.m.}$ in the
center of mass frame (the BSW effect) can be understood in terms of
three-dimensional velocities. Such an entity can be defined in the terad
formalism according to%
\begin{equation}
V^{(i)}=-\frac{e_{\mu}^{(i)}u^{\mu}}{e_{\mu}^{(0)}u^{\mu}},
\end{equation}
where $e_{\mu}^{(a)}$ ($a=0,1,2,3)$ $\ $is a tetrad attached to an observer.

The BSW effect occurs if the relative velocity of both particles $w$
approaches in this case that of light, so boost between them becomes
divergent. One of the easiest way to show this is to use the so-called OZAMO
frame for evaluation of components of $V^{(i)}$. This was shown in \cite{k},
where usual and critical particles were considered. Now, we can generalize the
corresonding results to scenarios with other types of particles. We can look
at the properties of $w$ in two ways. Either we (i) choose some a OZAMO frame
similarly to \cite{k} and calculate $w$ or (ii) carry out calculations of
velocity of particle 2 directly in the frame CO comoving with particle 1.

Let us consider both approaches separately.

\subsection{Evaluation of relative velocity in OZAMO frame}

At first, we discuss case (i). Then, the form of tetrads is given in Appendix
A. This approach has an additional motivation since the ZAMO frame (in both
versions - OZAMO and FZAMO) is a powerful standard tool for description of
particle motion near a black hole. It follows from (\ref{V}) that in the
horizon limit the ratio $V^{(2)}/V^{(1)}\rightarrow0$ as well as
$V^{(3)}/V^{(1)}$ and $V^{(1)}\rightarrow1$ for usual and subcritical
particles. Therefore, such kinds of particles hit the horizon
perpendicularily, so any two of them move in the same direction. Then, we
follow Sec. III of \cite{k}. Using the formulas of special relativity, we have
for the relative velocity of two particles%
\begin{equation}
w=\frac{V_{1}-V_{2}}{1-V_{1}V_{2}}\text{.}%
\end{equation}

Let near the horizon $V=1-\varepsilon$, where $\varepsilon\ll1$. Then, in the
limit under consideration%
\begin{equation}
w_{H}=\frac{\left\vert \varepsilon_{1}-\varepsilon_{2}\right\vert
}{\varepsilon_{1}+\varepsilon_{2}}.
\end{equation}

The key point here is the rate with which $V\rightarrow1$. For a usual
particles, it follows from (\ref{VNX}) that $\varepsilon_{1}\sim
\varepsilon_{2}\sim N^{2}$, so $w_{H}<1$ and the boost is finite.\ If particle
1 is usual and particle 2 is subcritical, $\varepsilon_{1}\sim N^{2}\sim
v^{p}$ and $\varepsilon_{2}\sim v^{p-2s_{1}}$, so $\varepsilon_{1}%
/\varepsilon_{2}\rightarrow0.$ Then, $w_{H}=1$ and the boost is infinite. If
both particles are subcritical, $\varepsilon_{1}\sim v^{p-2s_{1}}%
\sim\varepsilon_{2}$, $w_{H}<1$, so the boost is finite.

If particle 1 is critical or ultracritical, its limiting velocity $V_{1H}<1$
and it hits the horizon under some angle, not perpendicularly. Then, if
particle 2 is usual the relative velocity $w_{H}\rightarrow1$ and the boost is
infinite \cite{k}. The same is true if particle 2 is subcritical. If both
particles are critical or ultracritical, $w_{H}<1$ and the boost is finite.

\begin{table}[ptb]%
\begin{tabular}
[c]{|c|c|c|}\hline
Particle & Boost CO - OZAMO & Boost CO - FZAMO\\\hline
Usual & Infinite & Finite\\\hline
Subcritical & Infinite & Infinite\\\cline{1-1}\cline{2-3}%
Critical and ultracritical & Finite & Infinite\\\hline
\end{tabular}
\caption{Table showing different boost types between CO-OZAMO and CO-FZAMO
frames for different types of particles}%
\end{table}

We obtained the set of possibilities represented in Table II where we also
inserted the properties of the proper time necessary to reach the horizon. As,
in general, there are three relevant frames (OZAMO, FZAMO and CO), we present
the extended version of Table I in which we included charactersics in
corresponding frames in addition to OZAMO. In the last two column it is
implied that the CO is realized by a particle indicated in the first column.
In all cases FZAMO is realized by usual particles, while OZAMO is realized by
critical and ultracritical ones. In addition, the boost between FZAMO and
OZAMO is always infinite. This is in correspondence with row 4 in Table II in
\cite{gen} that shows that the relative gamma-factor diverges if one of
particles particle is usual (equivalent to FZAMO) and the second particle is
critical or ultracritical (equivalent to OZAMO). 

\subsection{Evaluation of relative velocity in CO frame}

Now, we turn to approach (ii). In doing so, we attach a tetrad to particle 1
and look at the velocity of particle 2 in this frame. We can obtain even more
detailed information concerning motion of particle than it was found above.
This is because now we are able to analyze not only the absolute value of
relative velocity but also its different components. To this end, we use the
results listed in Appendix A. We attach the tetrad $\widetilde{e}_{\mu}^{(i)}$
to some particle 1. Then, for any other particle 2 with four-velocity $\hat
{u}^{\mu}$ we have using (\ref{u}):%

\begin{equation}
\hat{V}^{i}=\frac{\widetilde{e}_{\mu}^{(i)}\hat{u}^{\mu}}{\gamma}\text{,}%
\end{equation}
where we use hat for quantities related to this particle, and $\gamma
=-\widetilde{e}_{\mu}^{(0)}\hat{u}^{\mu}$ is the Lorentz gamma factors of
relative motion of both particles. In other words, we measure the velocity of
particle 2 from the point of view of an observer comoving with particle 1.

Straightforward calculation gives us%
\begin{equation}
\hat{V}^{(1)}=\frac{1}{\gamma V}[-\sigma(X^{2}-N^{2})\frac{\hat{X}}{N^{2}%
X}+(\sigma L\frac{\hat{L}}{g_{\varphi}}+\sigma g_{\theta}u^{\theta}\hat{u}^{\theta
})+\frac{\hat{\sigma}P\hat{P}}{N^{2}}]\text{,} \label{v1}%
\end{equation}%
\begin{equation}
\hat{V}^{(2)}=\frac{\sqrt{g_{\theta}}}{P^{\prime}\gamma}(-\sigma\hat{\sigma
}\hat{P}u^{\theta}+P\hat{u}^{\theta})\text{,} \label{v2}%
\end{equation}%
\begin{equation}
\hat{V}^{(3)}=\frac{1}{P^{\prime}XV\sqrt{g_{\phi}}\gamma}[\hat{L}(P^{\prime
})^{2}-g_{\theta}u^{\theta}\hat{u}^{\theta}LN^{2}-\sigma\hat{\sigma}P\hat
{P}L]\text{,} \label{v3}%
\end{equation}
where%
\begin{equation}
\gamma=-\widetilde{e}_{\mu}^{(0)}\hat{u}^{\mu}=\frac{X\hat{X}-\sigma
\hat{\sigma}P\hat{P}-N^{2}g_{\theta}u^{\theta}\hat{u}^{\theta}}{N^{2}}%
-\frac{L\hat{L}}{g_{\phi}}\text{.}%
\end{equation}

If both particles coincide, one can check that (\ref{v0}) holds, so $\hat
{V}^{(i)}=0$ as it should be.

Now, in contrast to the OZAMO frame, a usual particle can hit the horizon
under an arbitrary angle and with an arbitrary value of velocity. For a
simplified case of motion of particle in the background of a spherically
symmetric static black hole this issue is discussed in Sec. 7 of \cite{flows}.
In this case, OZAMO turns into a static frame. It was stressed in \cite{flows}
that the boost between the static frame and CO is diverging near the horizon.
Then, if the angular component of velocity of particle 2 is very small in the
static frame, it becomes of order 1 in the frame comoving with particle 1,
provided particle 1 moves freely or under the action of finite force. Formulas
(\ref{v1}) - (\ref{v3}) can be of use in any analysis of particle motion near
the horizon in a physical (non-singular frame) analysis of particle motion,
say, in flows around a generic rotating black hole.

\section{Conclusions\label{concl}}

In this work we have investigated motion of particles moving near the horizon
of a generic axially symmetric black hole. In doing so, we did not constrain
consideration by motion in the equatorial plane. We also discussed different
types of horizon characterized by a set of integers $p$, $q$, $k$, $l$.
Moreover, we took into account the action of an arbitrary (regular) force on
such particles.

During the analysis of restrictions of the four-velocity caused by regularity
of forces in the properly chosen frame, we have found that the absolute value
of the projection of the 4-velocity on 2-sphere of constant radius (namely,
the quantity $s^{\mu},$ introduced in Sec. \ref{cons}) remains regular as one
approaches the horizon. Moreover, this statement is true for any type of
particle and is independent on whether the proper time for achieving the
horizon is regular or not. This statement fills the gap in previous
investigations where regularity of polar component of the four-velocity was
assumed but not proven. Also, we have to emphasize that this property does not
follow from kinematics and is obtained solely from the dynamics of considered
particles under the action of a regular force.

Concerning other restrictions, kinematics shows that the only new phenomenon
due to nonequatorial motion is the appearance of the so-called
\textquotedblleft belts\textquotedblright\ where the motion of particles may
occur. These \textquotedblleft belts\textquotedblright\ encircle the
equatorial plane and never reach poles $\theta=0,\pi.$ In particular, this
phenomenon reveals itself in the context of the BSW effect (see discussion in
\cite{kd} for the particular case of the Kerr metric). Other conditions come
mainly from the dynamics, specifying the range in which powers in the Taylor
expansion of the metric functions near the horizon may vary. The final list is
presented in Tables I and II. It is seen from them that a regular proper time
is possible only for usual particles (according to the classification,
introduced in \cite{gen}), while for subcritical, critical and ultracritical
ones the proper time diverges, provided the force is regular and the horizon
is regular.

As far as the properties of high-energy collisions are concerned,
classification of scenarios of the BSW effect found earlier for equatorial
motion is reproduced for nonequatorial one as well. Meanwhile, the results
obtained in the present work have quite general nature. They are not
restricted by high energy collisions and can apply to any physical or
astrophysical problem in which nonequatorial motion is essential. In
particular, this can be of use for description of particle motion and
collision in the accretion disc, propertis of chaotic motion, etc.

\section{Acknowledgement}

OZ was supported in part by the grants 2024/22940-0, 2021/10128-0 of S\~{a}o
Paulo Research Foundation (FAPESP).

\appendix

\section{Establishing comoving frame\label{comov}}

In this Appendix we describe how one can establish a comoving frame from the
OZAMO frame \cite{72}. We rely on relations (\ref{ozamo1}), (\ref{ozamo2}):
\begin{align}
e_{(0)}^{\mu}  &  =\dfrac{1}{N}(1,0,0,\omega),~~~e_{(1)}^{\mu}=\sqrt
{A}(0,1,0,0),\nonumber\\
e_{(2)}^{\mu}  &  =\dfrac{1}{\sqrt{g_{\theta\theta}}}(0,0,1,0),~~~e_{(3)}%
^{\mu}=\dfrac{1}{\sqrt{g_{\phi\phi}}}(0,0,0,1). \label{ozamo_fr}%
\end{align}

In this tetrad frame, the four-velocity reads%
\begin{equation}
u^{(a)}=\left(  \frac{X}{N},\sigma\frac{P}{N},\sqrt{g_{\theta}}u^{\theta
},\frac{L}{\sqrt{g_{\varphi}}}\right)  \text{.} \label{utet_2}%
\end{equation}

To establish transformations between OZAMO and CO frames, we have to find the
three-velocity in the OZAMO frame. Direct computation gives us%
\begin{equation}
V^{(i)}=-\frac{e_{\mu}^{(i)}u^{\mu}}{e_{\mu}^{(0)}u^{\mu}}=\left(
\frac{\sigma P}{X},\frac{\sqrt{g_{\theta}}u^{\theta}N}{X},\frac{LN}%
{\sqrt{g_{\varphi}}X}\right)  . \label{V}%
\end{equation}

The absolute value of the three-velocity is
\begin{equation}
|V|=\sqrt{1-\frac{N^{2}}{X^{2}}}. \label{VNX}%
\end{equation}

To transform from the OZAMO frame to the comoving frame, one has to perform
the following operations.

\begin{itemize}
\item Rotation in the $r\theta$ plane%
\begin{align}
e_{(1)}^{\prime}  &  =e_{(1)}\cos\delta+e_{(2)}\sin\delta,\text{ \ \ }%
e_{(0)}^{\prime}=e_{(0)},\label{1}\\
e_{(2)}^{\prime}  &  =e_{(2)}\cos\delta-e_{(1)}\sin\delta,\text{ \ \ }%
e_{(3)}^{\prime}=e_{(3)},
\end{align}
with
\begin{equation}
\tan\delta=\sigma\sqrt{g_{\theta}}\frac{u^{\theta}N}{P}.
\end{equation}

\item Rotation in the $r\varphi$ plane%
\begin{align}
e_{(1)}^{\prime\prime}  &  =e_{(1)}^{\prime}\cos\psi+e_{(3)}^{\prime}\sin
\psi,\text{ \ \ }e_{(0)}^{\prime\prime}=e_{(0)}^{\prime},\\
e_{(3)}^{\prime\prime}  &  =e_{(1)}^{\prime}\cos\psi-e_{(3)}^{\prime}\sin
\psi,\text{ \ \ }e_{(2)}^{\prime\prime}=e_{(2)}^{\prime},
\end{align}
with%
\begin{equation}
\tan\psi=\sigma\frac{LN}{\sqrt{g_{\varphi}}P^{\prime}},
\end{equation}
where%
\begin{equation}
P^{\prime}=\sqrt{X^{2}-N^{2}\left(  1+\frac{L^{2}}{g_{\varphi}}\right)  }.
\end{equation}

\item Perform a boost%
\begin{align}
\widetilde{e}_{(0)}  &  =\gamma(e_{(0)}^{\prime\prime}+\sigma|V|e_{(1)}%
^{\prime\prime}),\text{ \ \ }\widetilde{e}_{(2)}=e_{(2)}^{\prime\prime
},\label{b0}\\
\widetilde{e}_{(1)}  &  =\gamma(e_{(1)}^{\prime\prime}+\sigma|V|e_{(0)}%
^{\prime\prime}),\text{ \ \ }\widetilde{e}_{(3)}=e_{(3)}^{\prime\prime},
\label{b1}%
\end{align}
where $\gamma=\frac{1}{\sqrt{1-V^{2}}}=\frac{X}{N}.$
\end{itemize}

Combining all these transformations, one obtains that comoving tetrad is given
by%
\begin{equation}
\left(
\begin{array}
[c]{c}%
\widetilde{e}_{(0)}\\
\widetilde{e}_{(1)}\\
\widetilde{e}_{(2)}\\
\widetilde{e}_{(3)}%
\end{array}
\right)  =\left(
\begin{array}
[c]{cccc}%
\frac{X}{N} & \sigma\frac{P}{N} & \sqrt{g_{\theta}}u^{\theta} & \frac{L}%
{\sqrt{g_{\varphi}}}\\
\sigma\frac{X}{N}|V| & \frac{P}{N|V|} & \sigma\sqrt{g_{\theta}}\frac
{u^{\theta}}{|V|} & \sigma\frac{L}{\sqrt{g_{\varphi}}|V|}\\
0 & -\sigma\sqrt{g_{\theta}}\frac{u^{\theta}N}{P^{\prime}} & \frac
{P}{P^{\prime}} & 0\\
0 & -\sigma\frac{LN}{\sqrt{g_{\varphi}}X|V|}\frac{P}{P^{\prime}} &
-\sqrt{\frac{g_{\theta}}{g_{\varphi}}}\frac{Lu^{\theta}N^{2}}{P^{\prime}X|V|}
& \frac{P^{\prime}}{X|V|}%
\end{array}
\right)  \left(
\begin{array}
[c]{c}%
e_{(0)}\\
e_{(1)}\\
e_{(2)}\\
e_{(3)}%
\end{array}
\right)  , \label{table}%
\end{equation}
or, explicitly,%
\begin{align}
\widetilde{e}_{\mu}^{(0)}  &  =\left(  X+L\omega,-\sigma\frac{P}{\sqrt{A}%
N},-g_{\theta}u^{\theta},-L\right)  ,\\
\widetilde{e}_{\mu}^{(1)}  &  =\frac{1}{|V|}\left(  -\sigma\left(
X+L\omega-\frac{N^{2}}{X}\right)  ,\frac{P}{\sqrt{A}N},\sigma g_{\theta
}u^{\theta},\sigma L\right)  ,\\
\widetilde{e}_{\mu}^{(2)}  &  =\left(  0,-\sigma\frac{\sqrt{g_{\theta}%
}Nu^{\theta}}{\sqrt{A}P^{\prime}},\sqrt{g_{\theta}}\frac{P}{P^{\prime}%
},0\right)  ,\\
\widetilde{e}_{\mu}^{(3)}  &  =\frac{1}{|V|}\left(  -\sqrt{g_{\varphi}}%
\frac{P^{\prime}\omega}{X},-\sigma\frac{LN}{\sqrt{Ag_{\varphi}}X}\frac
{P}{P^{\prime}},-\frac{g_{\theta}}{\sqrt{g_{\varphi}}}\frac{u^{\theta}%
}{P^{\prime}}\frac{LN^{2}}{X},\sqrt{g_{\varphi}}\frac{P^{\prime}}{X}\right)  .
\end{align}

One can check that in this frame holds%
\begin{equation}
\widetilde{V}^{i}=-\frac{\widetilde{e}_{\mu}^{(i)}u^{\mu}}{\widetilde{e}_{\mu
}^{(0)}u^{\mu}}=(0,0,0)\text{,} \label{v0}%
\end{equation}
as it should be, and the four-velocity%
\begin{equation}
\widetilde{u}^{(a)}=\widetilde{e}_{\mu}^{(a)}u^{\mu}=(1,0,0,0).
\end{equation}

Calculation of acceleration in this frame $a_{C}^{(a)}=\widetilde{e}_{\mu
}^{(a)}a^{\mu}$ gives (\ref{af_0}-\ref{af_3}).

For completness and readers' convenience, we give below also the expressions
of the acceleration in the OZAMO frame $a_{O}^{(0)}=a^{\mu}e_{\mu(a)},$
although we do not use them directly:%
\begin{align}
a_{O}^{(0)}  &  =\sigma P\frac{\sqrt{A}}{N^{2}}(\partial_{r}X+L\partial
_{r}\omega)+\frac{u^{\theta}}{N}(\partial_{\theta}X+L\partial_{\theta}%
\omega),\label{ao_0}\\
a_{O}^{(1)}  &  =X\frac{\sqrt{A}}{N^{2}}\left(  \partial_{r}X+L\partial
_{r}\omega-\frac{N^{2}}{X}\frac{L\partial_{r}L}{g_{\varphi}}\right)
+u^{\theta}\sqrt{A}\left(  \sigma\partial_{\theta}\left(  \frac{P}{N\sqrt{A}%
}\right)  -\partial_{r}\left(  g_{\theta}u^{\theta}\right)  \right)
,\label{ao_1}\\
a_{O}^{(2)}  &  =\sigma\frac{\sqrt{A}}{N}\frac{P}{\sqrt{g_{\theta}}}%
\partial_{r}\left(  g_{\theta}u^{\theta}\right)  +u^{\theta}\partial_{\theta
}\left(  \sqrt{g_{\theta}}u^{\theta}\right)  +\frac{P^{2}\partial_{\theta}%
A}{2AN^{2}\sqrt{g_{\theta}}}+\nonumber\\
&  +\frac{X^{2}\partial_{\theta}N^{2}}{2N^{4}\sqrt{g_{\theta}}}+\frac
{LX\partial_{\theta}\omega}{\sqrt{g_{\theta}}N^{2}}-\frac{L^{2}\partial
_{\theta}g_{\varphi}}{2g_{\varphi}^{2}\sqrt{g_{\theta}}},\label{ao_2}\\
a_{O}^{(3)}  &  =\sigma\frac{\sqrt{A}}{N}\frac{P}{\sqrt{g_{\varphi}}}%
\partial_{r}L+\frac{u^{\theta}\partial_{\theta}L}{\sqrt{g_{\varphi}}}.
\label{ao_3}%
\end{align}

\section{Relation to Lorentz group transformations\label{ap}}

Above, we built the matrix that transforms a stationary frame to a frame that
is comoving with a particle. It includes dynamic characteristics of a particle
- such as energy and components of momentum. In this Appendix we give an
alternative approach that relies on the group transformations within the
Lorentz group.

It follows from eq. (\ref{table}) that%
\begin{align}
\tilde{e}_{(0)}  &  =\frac{X}{N^{2}}\partial_{t}+\sigma P\frac{\sqrt{A}}%
{N}\partial_{r}+u^{\theta}\partial_{\theta}+\left(  \frac{L}{g_{\varphi}%
}+\frac{X\omega}{N^{2}}\right)  \partial_{\varphi},\\
\tilde{e}_{(1)}  &  =\sigma\frac{X}{N^{2}}\partial_{t}+\frac{P}{|V|}%
\frac{\sqrt{A}}{N}\partial_{r}+\sigma\frac{u^{\theta}}{|V|}\partial_{\theta
}+\sigma\left(  \frac{L}{g_{\varphi}|V|}+\frac{X\omega}{N^{2}}\right)
\partial_{\varphi},\\
\tilde{e}_{(2)}  &  =\frac{-\sigma\sqrt{g_{\theta}}u^{\theta}N}{P^{\prime}%
}\sqrt{A}\partial_{r}+\frac{P}{P^{\prime}}\frac{1}{\sqrt{g_{\theta}}}%
\partial_{\theta},\\
\tilde{e}_{(3)}  &  =-\sigma\frac{LN}{\sqrt{g_{\phi}}XV}\frac{P}{P^{\prime}%
}\sqrt{A}\partial_{r}-\frac{Lu^{\theta}N^{2}}{\sqrt{g_{\phi}}P^{\prime}%
XV}\partial_{\theta}+\frac{P^{\prime}}{VX\sqrt{g_{\phi}}}\partial_{\varphi}.
\end{align}

Let us introduce a new parameter $q$ and use also angles $\delta$ and $\psi,$
satisfying:%
\begin{align}
\cosh q  &  =\frac{X}{N},\text{ \ \ }\sinh q=\sigma\frac{X|V|}{N},\\
\cos\delta &  =\frac{P^{\prime}}{X|V|},\text{ \ \ }\sin\delta=\sigma\frac
{LN}{\sqrt{g_{\varphi}}X|V|},\\
\cos\psi &  =\frac{P}{P^{\prime}},\text{ \ \ }\sin\psi=\sigma\sqrt{g_{\theta}%
}\frac{u^{\theta}N}{P^{\prime}}.
\end{align}

Then,
\begin{equation}
\sigma\frac{P}{N}=\sinh q\cos\delta\cos\psi\text{, }\frac{L}{\sqrt{g_{\phi}}%
}=\sin\delta\sinh q\text{, }\sqrt{g_{\theta}}u^{\theta}=\sinh q\cos\delta
\sin\psi\text{,}%
\end{equation}

and%
\begin{align}
\tilde{e}_{(0)}  &  =\frac{\cosh q}{N}\partial_{t}+\sinh q\cos\delta\cos
\psi\sqrt{A}\partial_{r}+\\
&  +\sinh q\cos\delta\sin\psi\frac{1}{\sqrt{g_{\theta\theta}}}\partial
_{\theta}+(\frac{\sinh q\sin\delta}{\sqrt{g_{\phi}}}+\frac{\omega\cosh q}%
{N})\partial_{\varphi} ,\nonumber\label{0}\\
\tilde{e}_{(1)}  &  =\frac{1}{N}\sinh q\partial_{t}+\cosh q\cos\delta\cos
\psi\sqrt{A}\partial_{r}+\nonumber\\
&  +\frac{\cosh q\cos\delta\sin\psi}{\sqrt{g_{\theta}}}\partial_{\theta
}+(\frac{\cosh q\sin\delta}{\sqrt{g_{\phi}}}+\frac{\omega\sinh q}{N}%
)\partial_{\varphi},\\
\tilde{e}_{(2)}  &  =-\sin\psi\sqrt{A}\partial_{r}+\cos\psi\partial_{\theta
},\\
\tilde{e}_{(3)}  &  =-\sin\delta\cos\psi\text{ }\partial_{r}-\sin\psi
\sin\delta\text{ }\partial_{\theta}+\cos\delta\text{ }\partial_{\varphi}.
\label{3}%
\end{align}

The table that enters eq. (\ref{table}) can be rewritten in the form%

\begin{equation}
\Lambda=\left(
\begin{array}
[c]{cccc}%
\cosh q & \sinh q\cos\delta\cos\psi & \sinh q\cos\delta\sin\psi & \sinh
q\sin\delta\\
\sinh q & \cosh q\cos\delta\cos\psi & \cosh q\cos\delta\sin\psi & \cosh
q\sin\delta\\
0 & -\sin\psi & \cos\psi & 0\\
0 & -\sin\delta\cos\psi & -\sin\psi\sin\delta & \cos\delta
\end{array}
\right)  .
\end{equation}

One can check that in the equatorial case ($\theta=\frac{\pi}{2}$) this
parametrization reduces to that in \cite{obs} and eq. (42) of \cite{tz}.

\subsection{Null tetrad}

We also use the null tetrad bearing in mind possible application in the
framework of the Newman-Penrose formalism.

Let us introduce the null vectors $k$, $l$, \thinspace$m$, $\bar{m}$ according
to%
\begin{align}
e_{(0)}  &  =\frac{k+l}{\sqrt{2}},\text{ \ \ \ \ \ }e_{(1)}=\frac{k-l}%
{\sqrt{2}},\label{e}\\
e_{(2)}  &  =\frac{m+\bar{m}}{\sqrt{2}}\text{, \ \ }e_{(3)}=\frac{m-\bar{m}%
}{\sqrt{2i}}\text{.} \label{m}%
\end{align}

Then, one check that the transformations (\ref{1}-\ref{b1}) with $v=\tanh q$
can be rewritten in the form%
\begin{align}
\tilde{k}  &  =e^{q}\left[  k\left(  \frac{1+\cos\delta\cos\psi}{2}\right)
+l\left(  \frac{1-\cos\delta\cos\psi}{2}\right)  +\right. \nonumber\\
&  \left.  +m\left(  \frac{\sin\psi\cos\delta}{2}+\frac{\sin\delta}%
{2i}\right)  +\bar{m}\left(  \frac{\sin\psi\cos\delta}{2}-\frac{\sin\delta
}{2i}\right)  \right]  ,\\
\widetilde{l}  &  =e^{-q}\left[  l\left(  \frac{1+\cos\delta\cos\psi}%
{2}\right)  +k\left(  \frac{1-\cos\delta\cos\psi}{2}\right)  \right.
-\nonumber\\
&  \left.  -m\left(  \frac{\sin\psi\cos\delta}{2}+\frac{\sin\delta}%
{2i}\right)  -\bar{m}\left(  \frac{\sin\psi\cos\delta}{2}-\frac{\sin\delta
}{2i}\right)  \right]  ,\\
\tilde{m}  &  =m\left(  \frac{\cos\psi+\cos\delta-i\sin\psi\sin\delta}%
{2}\right)  +\overline{m}\left(  \frac{\cos\psi-\cos\delta-i\sin\psi\sin
\delta}{2}\right)  -\nonumber\\
&  -(k-l)\frac{\sin\psi+i\sin\delta\cos\psi}{2},\\
\widetilde{\overline{m}}  &  =m\left(  \frac{\cos\psi-\cos\delta+i\sin\psi
\sin\delta}{2}\right)  +\overline{m}\left(  \frac{\cos\psi+\cos\delta
+i\sin\psi\sin\delta}{2}\right) \nonumber\\
&  -(k-l)\frac{\sin\psi-i\sin\delta\cos\psi}{2}.
\end{align}

If one use the vectors $\tilde{e}$ constructed from tilted null tetrad similar
to (\ref{e}), (\ref{m}), the result will coincide with (\ref{0}-\ref{3}).

The above formulas establish relationship between the Lorentz transformations
in a locally flat space-time and particle dynamics in the curved background.

\end{document}